\documentclass[preprint]{aastex}

\shorttitle{Mira C and O Isotopes}
\shortauthors{Hinkle et al.}

\begin{document}

\title{CARBON AND OXYGEN ISOTOPIC RATIOS FOR NEARBY MIRAS}

\author{KENNETH H. HINKLE}
\affil{National Optical Astronomy Observatory\\
P.O. Box 26732, Tucson, AZ 85726 USA}
\email{khinkle@noao.edu}

\and

\author{THOMAS LEBZELTER}
\affil{Department of Astrophysics, University of Vienna\\ 
T{\"u}rkenschanzstrasse 17, 1180 Vienna, Austria}
\email{thomas.lebzelter@univie.ac.at}

\and

\author{OSCAR STRANIERO}
\affil{INAF, Osservatorio Astronomico di Collurania\\ 
I-64100 Teramo, Italy, and INFN-LNGS, Assergi (AQ), Italy}
\email{straniero@oa-teramo.inaf.it}

\begin{abstract}
Carbon and oxygen isotopic ratios are reported for a sample of 46
Mira and SRa-type variable AGB stars.  Vibration-rotation first and
second overtone CO lines in 1.5 to 2.5 $\mu$m spectra were measured
to derive isotopic ratios for $^{12}$C/$^{13}$C, $^{16}$O/$^{17}$O,
and $^{16}$O/$^{18}$O.  Comparisons with  previous measurements
for individual stars and with various samples of evolved stars, as
available in the extant literature, are discussed.  Models for solar
composition AGB stars of different initial masses are used  to
interpret our results.  We find that the majority of the M stars
had main sequence masses $\le 2$ M$_\odot$ and have not experienced
sizable third dredge-up episodes. The progenitors of the four
S-type stars in our sample are slightly more massive.  Of the six
C stars in the sample three have clear evidence relating their
origin to the occurrence of the third dredge-up.  Comparisons with
O-rich presolar grains from AGB stars that lived before the formation
of the solar system reveal variations in the interstellar medium
chemical composition.  The present generation of low-mass AGB stars,
as represented by our sample of Long Period Variables, show a large
spread of $^{16}$O/$^{17}$O ratios, similar to that of group 1
presolar grains and in agreement with theoretical expectations for
the composition of mass 1.2 to 2 M$_\odot$ stars after the first
dredge up.  On the contrary, the $^{16}$O/$^{18}$O ratios of
present-day Long Period Variables are definitely smaller than those
of group 1 grains. This occurrence is most probably a consequence
of the the decrease with time of the $^{16}$O/$^{18}$O ratio in the
interstellar medium due to the chemical evolution of the Milky Way.
One star in our sample has an O composition similar to that of group
2 presolar grains originating in an AGB star undergoing extramixing.
This occurrence may indicate that the extramixing process is hampered
at high metallicity or, equivalently, favored at low metallicity.
Similar to O-rich grains no star in our sample shows evidence of
hot bottom burning, expected for massive AGB stars.

\end{abstract}


\section{INTRODUCTION}

Mass is returned from stars to the ISM through two primary routes,
supernovae explosions of massive stars and the asymptotic giant
branch (AGB) stellar wind of low- and intermediate-mass stars \citep{sedlmayr_1994}.
The percentages contributed through the various evolutionary paths
depend on multiple factors including metallicity but it is clear
that by early in the evolution of the universe material enriched
through AGB processing is present \citep[see review
by][]{maio_tescari_2015}.  The majority of the mass return from
low-mass stars to the ISM occurs in the terminal AGB stages when
large amplitude pulsation plays a critical role
\citep{vassiliadis_wood_1993}.  The composition of the AGB material
ejected depends on the initial stellar composition, nuclear processing,
and mixing of the stellar material over the lifetime of the star.
Of special interest in tracking these processes are the isotopic
ratios of the light elements carbon and oxygen. There are two stable
isotopes of carbon, $^{12}$C and $^{13}$C, and three of oxygen,
$^{16}$O, $^{17}$O, and $^{18}$O.

The atmospheric composition of the AGB stars, in addition to being
important in understanding the evolution of elemental abundances, is also important
in the evolution of galaxies.  Abundances originating on the AGB play a key factor in
the condensation of circumstellar dust around these stars.  The  
dust from AGB stars is in turn the origin of a large fraction of interstellar grain nuclei
\citep{gehrz_1989}.  
Interstellar grains from the protosolar nebula have been conserved
in meteorites.  These dust particles can be dissected out and the
isotopic ratios of carbon and oxygen measured in the lab \citep{hoppe10}.
This allows a comparison of abundances from interstellar dust with
abundances measured in the atmospheres of AGB stars.

The surface carbon and oxygen isotopic ratios found in AGB stars
are modified by CNO cycle and He burning and mixing processes that
have taken place in the stellar interior over the lifetime of the
star \citep[e.g.,][]{dearborn92}.  In a region where an incomplete
CNO cycle takes place $^{13}$C and $^{17}$O increase whereas
$^{12}$C and $^{18}$O are depleted. In the heart of an
H-burning zone, where the temperature is higher and the CNO cycle
attains the equilibrium, $^{13}$C and $^{17}$O are also consumed.
On the other hand, $^{12}$C is the main product of He burning
through the well-known $3\alpha$ reaction.  $^{18}$O can be produce
in a region where the He burning is active through $\alpha$ capture
on $^{14}$N followed by the $^{18}$F $\beta$-decay.  The ashes of
these fundamental nucleosynthesis processes can appear at the surface
as a consequence of a deep mixing.  These `dredge-up' events
produce substantial modification of the atmospheric abundances of
carbon and oxygen isotopes.

The first dredge-up (FDU) occurs in all stars of nearly solar
metallicity, independent of mass, when the star leaves the main
sequence and approaches the red giant branch (RGB). 
The second dredge-up (SDU) takes place during the early-AGB phase
but only for stars of $> 4$ $M_\odot$. In both the FDU
and the SDU material previously exposed to the CNO cycle is mixed
and brought to the surface.  A third dredge-up (TDU)
occurs during the late part of the AGB phase. Recurring
thermal pulses during the AGB phase caused by He-burning runaways push the external
envelope of the star outward and it expands and cools. As a result
shell-H burning ceases and external convective instability penetrates
into the H-exhausted core. Products of He burning
are mixed to the stellar surface. As a general rule the stronger the
thermal pulse the deeper the following TDU episode. 
Models suggest that deep TDU episodes occur in the stellar mass range 1.5 $M_\odot$ to 3 $M_\odot$.
In any case, stars less massive than about 1.2 -- 1.3 $M_\odot$ \footnote{The limit depends on
the pre-AGB mass loss and on the initial chemical composition.} arrive on the AGB with too
small an envelope and do not experience TDU.
Thermal pulses are weaker in massive AGB stars so these stars
experience shallower TDUs \citep[for a review see][]{iben_renzini1983,straniero2006}.

For the most massive AGB stars ($M > 6~M_\odot$) another
nucleosynthesis process affects the composition, hot bottom burning
(HBB).  In this case, the temperature at the base of the 
convective envelope that is always present is large enough
to activate the CNO cycle and, in the most extreme cases, the Ne-Na
and the Mg-Al cycles \citep[see e.g.][]{doherty2014}.  This
coupling of nuclear burning and convective mixing causes important
changes of the atmospheric abundances of light isotopes.

In addition to thermal convection other processes can be responsible
for the mixing of the material modified by the internal nuclear
burning. Among these processes are rotational induced (meridional)
mixing, magnetic buoyancy, and thermohaline circulation
\citep{charbonnel_lagarde_2010,nordhaus2008}.  The effective
importance of these phenomena in the modification of stellar chemical
patterns is highly debated \citep{karakas_et_al_2010, busso_et_al_2010, 
maiorca_et_al_2012}.  They are generically called extra-mixing
processes regardless of the true nature of the underlying physical
mechanism.  Nevertheless, there is clear proof of the existence of
one (or more) of these phenomena.  When low-mass stars ($M$~\textless~$2~M_\odot$) 
climb the brightest portion of the RGB  $^{12}$C/$^{13}$C is substantially reduced
\citep{gilroy_89,gilroy_brown_91,gratton_et_al00,abia_2012}.  For
solar metallicity stars this reduction may be up to 50\% of the
value attained after the first dredge-up.  Similar evidence is also
found in some low-mass AGB stars \citep{lebzelter_et_al08} but it
does not seem a general rule \citep{lederer_et_al09}.

Not surprisingly it also has been demonstrated that initial
isotopic compositions differ between stars.  In a previous article we studied the
oxygen isotopic composition of a few giant stars belonging to open
clusters of known age with distance from the galactic center between
8 and 10\,kpc, similar to that of the sun.  We found that the initial
oxygen isotopic ratios of these open clusters were different from
those measured in the solar system \citep{lebzelter_et_al15}.  This
presents a further complication in the interpretation of the observed
isotopic abundances in terms of physical phenomena taking place
during the stellar lifetime.

In this paper, we present new measurements of carbon and oxygen
isotopic ratios in a large sample of long period variables (LPVs).
Our
sample is limited to large amplitude LPVs of the Mira and SRa
variable classes clearly placing these objects on the AGB \citep{wood_2015}.  The 
Mira LPVs are the most luminous of the LPV
family of variables and represent the fundamental mode pulsators
\citep{wood_2015}.  Mass-loss rate increases exponentially with
pulsation period for Miras with period below $\sim$800 days
\citep{vassiliadis_wood_1993}. At extremely long period the stars
become obscured by circumstellar dust.  Our sample includes only
unobscured stars with period \textless\,600 days.

Abundance measurements in LPVs can be directly compared
with both theoretical expectations for AGB models and abundances
measured in presolar grains.  However, LPV spectra are
notoriously difficult to analyze.  The optical spectrum is dominated
by scattering processes \citep{ireland_et_al_2008}.  The photospheric
lines, which can be detected only in the infrared, undergo large
temperature and velocity variations \citep{hinkle_et_al_1982}.
Model atmospheres and spectrum synthesis techniques are clearly
limited in dealing with these spectra \citep{mcsaveny07,lebzelter_et_al14}.
Here we employ an empirical technique that utilizes broad-band
near-IR spectra and is best suited for the analysis of isotopomers.
In the following section we will discuss the analysis of the spectra
and then discuss the resulting isotopic ratios.

\section{OBSERVATIONS AND REDUCTIONS}

The observational material in this paper is drawn entirely from the
archives of the Kitt Peak National Observatory (KPNO) 4-meter Mayall
telescope Fourier Transform Spectrometer (FTS).  All the program
stars are AGB variables, restricted to those that are Mira and SRa
variables.  We differentiated between the various classes of SR
type variables using the latest General Catalog of Variable Stars
(GCVS).  In a few cases we have included variables with GCVS
classification of SRb but with an amplitude of an SRa.   The complete
set of spectra used for this project is listed in Table \ref{t:obslog}.

The FTS was operated between the years 1975 and 1995 by KPNO as a
coude instrument on the Mayall Telescope \citep{hall_et_al_1979}.
FTS technology provides high throughput but lacks multiplex advantage
\citep{ridgway_hinkle_1992} and as a result is suitable only for
observations of bright objects.  The redeeming features of FTS
include a large number of desirable instrumental characteristics
such as very large wavelength coverage, no scattered light, and
well defined instrumental profile.  Hence the wide application of
this technique in laboratories but not in ground-based night-time
astronomy.  This said, over the lifetime of the Mayall FTS thousands
of bright stars in the northern sky were observed.  Many of the
brightest stars in the infrared sky are variable AGB stars and a
rich selection of Mira spectra were archived.  Many of these are
in time series.

All the program stars were observed in a broad band-pass mode
covering the 1.5 to 2.5\,$\mu$m spectral region.   The band-pass
filters employed were a green glass filter to block long wavelengths
and a 1.6\,$\mu$m interference long-pass filter.  Both filters were
wedged and the combination is free of fringing.  The transmission
curve is smooth and featureless except for a 10\% absorption in the
glass from about 2.2 to 2.3\,$\mu$m.  The smoothness of the filter
transmission and very large wavelength coverage are useful for
estimating the continuum.  While the band pass and sampling
characteristics of the spectra became largely uniform after the
commissioning of the facility instrument in 1979, the broad spectral
response depended on the beamsplitter coatings.  Due to required
maintenance for exposed optical surfaces the coatings were replaced
several times over the lifetime of the instrument.  In analyzing
the data for this paper we found that optimizing these coatings for
the much requested $K$ band resulted in decrease in the bluest $H$
band response by about a factor of two after 1982.

Rather than wavelength the natural unit of an FTS is inverse
wavelength, i.e. wavenumber ($\sigma$). Following standard convention we use cm$^{-1}$ units.  The
spectral sampling ($\Delta\sigma$) is constant in
wavenumber and typically $\sim$0.070 cm$^{-1}$ for this data set.
This corresponds to spectral resolution (R = $\sigma/\Delta\sigma$) 
$\sim$ 57000 at the red end (4000 cm$^{-1} =
2.5~\mu$m) of the spectrum and R$\sim$96000 at the blue end (6670
cm$^{-1} = 1.5~\mu$m).  A few spectra of slightly lower resolution
are used as well. The lowest resolution used in this paper is 0.15
cm$^{-1}$ which corresponds to R=27000 at the red end of the spectrum.
FTS spectra are typically apodized to dampen the side lobes of the
sinc function instrumental profile.  Damping the amplitude of
the instrumental profile side lobes lowers the resolution and results
in a corresponding increase in signal-to-noise.  All spectra taken
after about 1979 have been apodized by function I2 of
\citet{norton_beer_1976}.  Prior to 1979 a prototype FTS was employed
that had path-dependent vignetting.  This resulted in an instrumental
apodization and we have not apodized these spectra.  The choice of
apodizing function is shown in the Table of Observations
(Table\,\ref{t:results}).

The limiting $K$-band magnitude for the broadband observations
discussed in this paper was about +1.  At this limiting magnitude
the observation of hot stars for use as telluric standards is very
time consuming.  Most of the program stars were short observations,
frequently filling a short time slot in another program, and
telluric reference spectra were not observed.  The measurement
technique depends on observing many lines and those lines blended
with telluric features are easily discarded.  Hence no attempt has
been made to correct the spectra for telluric absorption using, for
instance, telluric synthesis techniques.

Visual phases have been computed for the sample stars where data
from the AAVSO were available. Wherever possible the maximum directly
preceding our observation was used as a reference point.  In a few
cases the time of the maximum could not be determined accurately
due to the lack of sufficient data. These cases are marked with a
colon in Table\,\ref{t:results}.  Periods were taken from the GCVS.
For stars where we found additional period determinations in the
literature we determined the period at the time of the observation
by measuring the time difference to the next maximum.

The spectra were taken for other projects, mainly to monitor
kinematics.  Many spectra have previously been discussed in the
literature \citep{hinkle_et_al_1982, hinkle_et_al_1984, hinkle_et_al_1997,
hinkle_et_al_2000, lebzelter_et_al_1999}.  Where multiple spectra
of a star exist we have, when possible, selected spectra with the
best signal-to-noise and phases that are post-maximum.  Work on
stellar kinematics shows that the photospheric spectrum is simplest
and richest a few tenths of a period after maximum. In a few cases
we have analyzed multiple phases. For some stars only one spectrum
is available and selection of phase is not possible.

\section{ANALYSIS}

\subsection{Curve of Growth Analysis}

The stellar sample discussed here consists entirely of large amplitude
AGB variables, Miras and SRa variables. As mentioned above spectrum synthesis for these
objects continues to be beyond the grasp of model atmospheres and
line modeling techniques \citep{lebzelter_et_al14}.  To undertake
a first analysis of isotopic ratios in these stars we have turned
to a largely empirical technique.  The first and second overtone
CO lines in the 2.3 $-$ 2.5\,$\mu$m and 1.5 $-$ 1.8 $\mu$m regions,
respectively, were employed.  As noted above the observed spectra
cover the entire 1.5 $-$ 2.5 $\mu$m region allowing both overtones
to be analyzed in a single spectrum.  The basic reduction techniques
have been discussed in \citet{hinkle_et_al_1976}, \citet{hinkle_1978}, and \citet{hinkle_et_al_1982}.
A similar technique for measuring isotopic ratios can also be found
in \citet{dominy_wallerstein87}.  These papers employ curve of
growth (CoG) analysis.  CoG is a classical technique for deriving
elemental abundances that includes many simplifying assumptions
that can be dealt with more thoroughly through spectrum synthesis.
However, CoG is a powerful technique if only isotopic ratios are
derived.  The comparison of curves of growth for isotopically
substituted molecules remove many of the uncertainties associated
with the atmospheric structure. We briefly describe our approach
in this section.

As a first step, the continuum was set by using high points across
the entire band pass.  The technique typically works well owing to
a featureless filter transmission function with gentle curvature
at the edges.  However, Mira variables have emission lines at
pre-maximum phases \citep{hinkle_barnes_1979, hinkle_et_al_1982}
that can result in a high continuum placement.  When possible in this analysis we
have not used phases when emission lines can be present, between
roughly 0.8 and 0.0.  For some objects we had no choice but to use 
spectra in this range of phases.  In these cases we examined the high 
points at high resolution to attempt to determine if they were emission 
lines.

\begin{figure}
\epsscale{0.8}
\plotone{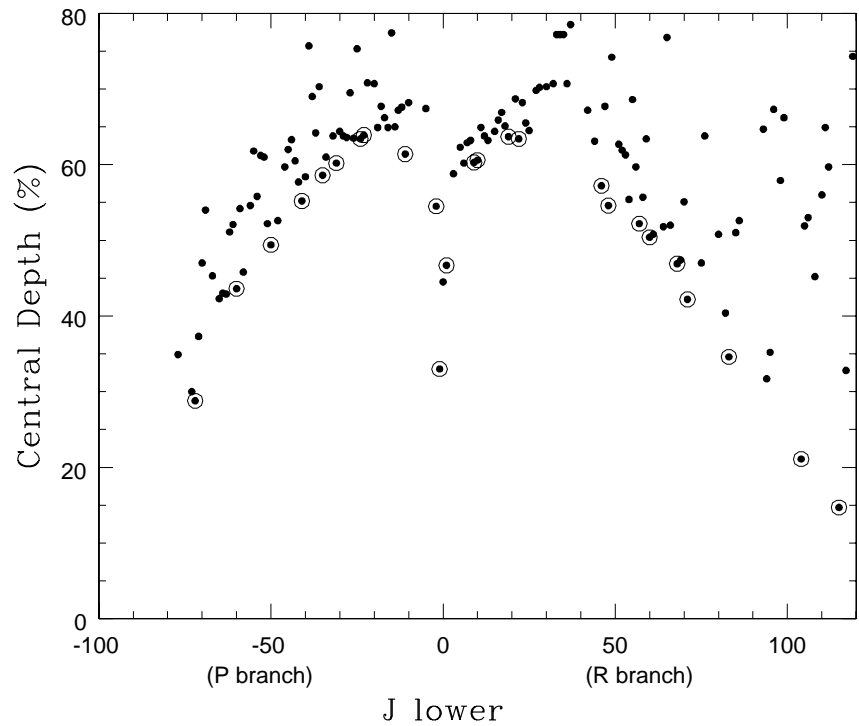}
\caption{Central depth measurements for the P and R branches of the
$^{12}$C$^{16}$O 4-1 band in T Cep 77 Nov 8.  The J'' values for
the P branch are shown as negative values.  The circled points were
selected as least blended and used in the curve of growth analysis.
\label{f:depths_vs_j}
}
\end{figure}

The velocity of the most abundant CO isotopologue, $^{12}$C$^{16}$O,
was determined from the second overtone lines using cross correlation
against laboratory line positions. This velocity was used
to set the apparent wavelengths of all first and second overtone
CO transitions in the filter band pass.  FTS spectra have an intrinsically accurate wavenumber scale 
and do not depend on a dispersion relation.  Our technique makes
use of accurate wavenumber scale.  A line list was generated
covering the vibration-rotation lines expected under LTE in cool
(2000 $-$ 4000 K) star atmospheres.  The depths of all spectral lines
at the line list positions were measured.  Plots of line central
depth versus rotational quantum number were produced for each
vibration transition.  The depths form a conspicuous lower bound
with a distinctive shape (Figure \ref{f:depths_vs_j}) resulting
from the rotational population \citep{herzberg_1950}.  A line list
was produced from the lines at the lower bound.  The equivalent
widths of the lines in the list were measured and a calibration of
central depth to equivalent width produced.  To check on the validity
of the selected lines we compared the lists for the program stars.
Lists of lines for the various CO isotopologues that are in common to 
at least four stellar spectra are given in the appendices.

The analysis proceeds by assuming a Planckian source function and
an isothermal atmosphere.  The isothermal atmosphere, while a very
crude assumption, reflects the excitation temperature of the molecular
line forming region.  With these assumptions there is a simple
relation between the strength of the spectral line,  the excitation
energy of the line and the excitation temperature of the gas in the
stellar atmosphere \citep{hinkle_et_al_1976}.  By minimizing the
separation in the curves of growth between high and low excitation
lines a value for the excitation temperature can be established.
The excitation temperature was set from the rich selection of lines
in the $^{12}$C$^{16}$O second overtone and applied to all the
overtones and isotopes.  The isotopic ratio is measured from the
shift between the different curves (Figure ~\ref{f:exampleCoG}).

The use of one molecule combined with comparing lines of similar strength and
excitation eliminates many complications.  In a typical case
$^{12}$C/$^{13}$C is determined from CO second overtone lines
spanning a large range of vibrational transitions with $^{13}$CO
mid-excitation lines having the same depth as $^{12}$CO low and high
excitation lines.  As a result the isotopic ratio will be temperature
sensitive.  On the other hand the $^{17}$O isotope was measured
using mainly the lines 2-0 C$^{17}$O R 25 to 33.  In this case only
a small range of excitation energy is probed.  The line depths are
typically similar to 3-0 $^{12}$C$^{16}$O lines that have similar
excitation energy so there is little temperature sensitivity.  As
part of the evaluation of the uncertainties the excitation temperature
was varied through a reasonable range.

\subsection{Isotopic ratios}

\begin{figure} 
\epsscale{0.7}
\plotone{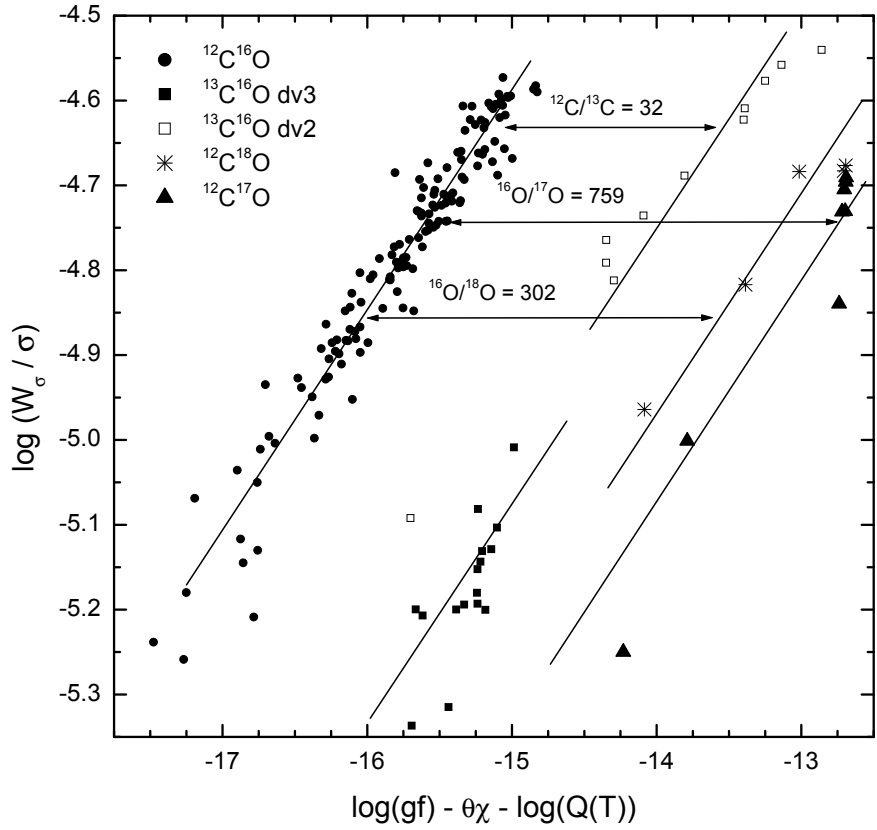}
\caption{Curves of growth for near-infrared first- and second-overtone
lines derived from lines of various CO isotopologues measured in
the 1977 Nov 8 spectrum of T Cep.  The offsets of the various curves
along the x-axis are shown by arrows and the corresponding isotopic
ratios are given. The excitation temperature has been set to
2900\,K.\label{f:exampleCoG}} \end{figure}

As described in the previous paragraphs, the isotopic ratio is
determined from the shift in the abscissa between the two CoG, e.g.,
one for $^{12}$CO and one for the rarer isotope.  An illustration
of this method is given in Figure~\ref{f:exampleCoG} for the Mira
T Cep.  The reduced equivalent width, W$_\sigma/\sigma$, is shown
as a function of the LTE line strength, $( g f / Q(T) ) 10^{-\Theta\chi}$,
where W$_\sigma$ the equivalent width in wavenumbers, $\sigma$ is
the wavenumber of the transition, g is the statistical weight (for
vibration-rotation CO g=$2 \rm{J} ''\!+\!1$), f is the oscillator
strength, $\Theta$ is 5040 divided by the excitation temperature,
$\chi$ is the energy of the lower level of the transition, and Q(T)
is the partition function that includes the relative number of
particles of the studied species.  The CO  wavenumbers and excitation
energies were computed from \citet{george_et_al_1994}.  The oscillator
strengths are from \citet{goorvitch_1994a, goorvitch_1994b}.  Isotopic
ratios determined for the program stars are listed in Table
\ref{t:results}.

The CoG shown in Figure~\ref{f:exampleCoG} is also typical in the
sense that the lower left end of the CoG is less clearly defined
with larger scatter resulting from a relatively larger noise
contribution to the weaker lines. In some cases we see a change in
the slope as the lines transition to the weak line portion of the
CoG and also possibly have a larger excitation temperature. Whenever
possible we tried to avoid using the weakest lines or gave these
lines a lower weight.  Of course in some cases isotopic lines are
only represented by weak lines and this is reflected in the
uncertainties.

For all stars of our sample the number of second overtone $^{12}$CO
lines was sufficiently large to define the CoG very well.  Typically
first overtone $^{12}$CO lines suitable for abundance analysis are
poorly represented.  Only the 2-0 through 5-3
bands can be observed before the spectrum is cut off by telluric absorption
near 2.5 $\mu$m. In Miras and SRa stars of near-solar abundance the
first overtone $^{12}$CO 2-0 lines below J$\sim$50 are strongly
saturated and lines up to J$\sim$30 are typically blended with a
line component formed in the extended atmosphere \citep{hinkle_et_al_1982}.
The selection of $^{13}$CO lines usually permitted a determination
of the $^{12}$C/$^{13}$C ratio from both the first and second
overtone lines.  The $^{12}$C/$^{13}$C value was then determined
by computing the average of both measurements.  The small difference
found between the two results supports the reliability of our
measurements.

\begin{figure}
\epsscale{1.1}
\plotone{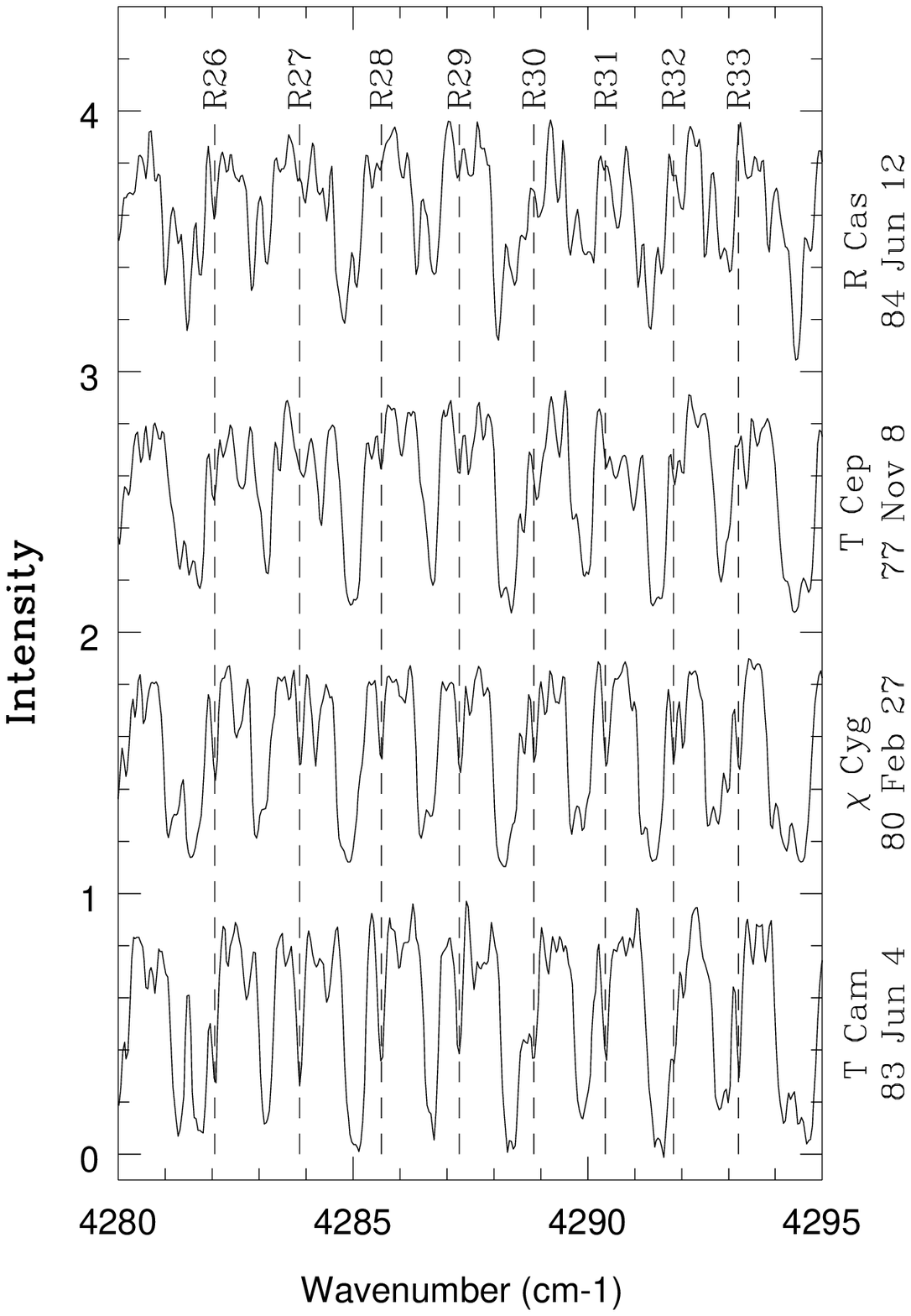}
\caption{R branch $^{12}$C$^{17}$O 2-0 lines identified in
the spectra of four Miras.  The selected Miras show a large range in $^{12}$C$^{17}$O
lines strengths.  The spectra are shown normalized to the continuum
with the $^{12}$C$^{17}$O lines at the laboratory frequencies.  The
$^{12}$C$^{17}$O 2-0 R branch lines are well placed in the spectrum
and identifications are secure. The strong features are low excitation
$^{12}$C$^{16}$O lines.
\label{f:fig12C17O} }
\end{figure}

\begin{figure}
\epsscale{1.1}
\plotone{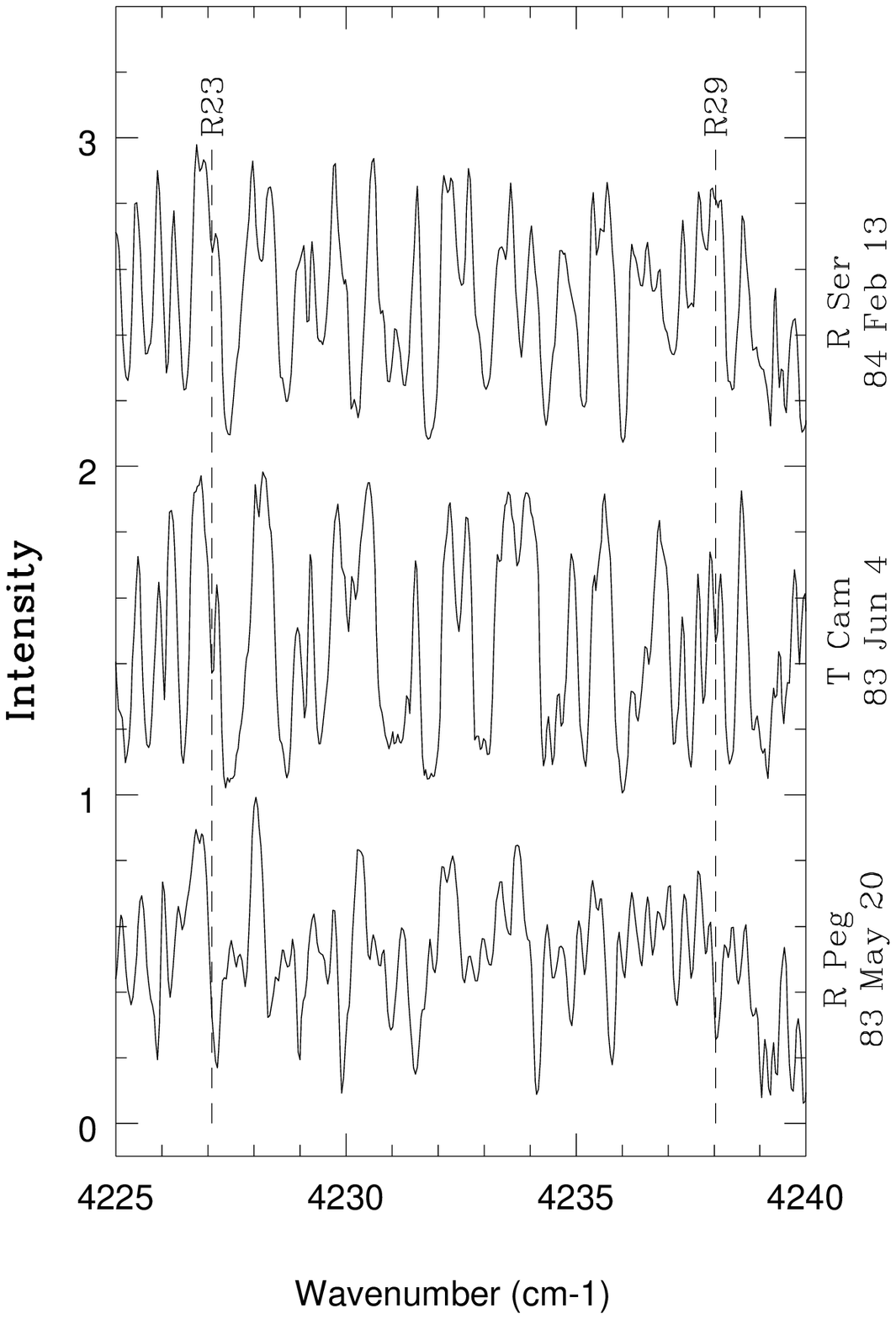}
\caption{Two $^{12}$C$^{17}$O 2-0 R branch lines, R23 and R29, identified in the spectra 
of three Miras.  The selected Miras show a large range in
$^{12}$C$^{18}$O lines strengths.  The spectra are shown normalized
to the continuum with the $^{12}$C$^{18}$O line at the laboratory
frequencies.  The $^{12}$C$^{18}$O lines are poorly placed in the
spectrum.  The isotopic ratio generally depends on a meager selection
of lines typically including either 2-0 R29 or R45. R23 which has
been used by other investigators in non-variable stars, when not an obvious blend, appears
slighter stronger than R29 perhaps because of an unidentified blend.
\label{f:fig12C18O} } 
\end{figure}

The $^{16}$O/$^{17}$O ratios are based on at least three and typically
five or more C$^{17}$O lines.  A section of the $^{12}$C$^{17}$O
2-0 band containing some of the strongest lines is well placed
between other CO lines (Figure \ref{f:fig12C17O}) and we feel the
measurements of these lines are robust.  For the $^{16}$O/$^{18}$O
ratios, the number of useable lines was smaller, and for several
stars only one line was available (Figure \ref{f:fig12C18O}). In
these cases, which are marked in Table \ref{t:results}, the shift
on the $^{12}$CO CoG could not be confirmed by a second line, so
that the $^{16}$O/$^{18}$O ratios have to be taken with caution.
The line used by \citet{lebzelter_et_al15}, 2-0 R23, frequently
appeared slightly too strong in the Mira spectra (Figure
\ref{f:fig12C18O}).  The $^{16}$O/$^{18}$O ratios are clearly more
uncertain than the $^{12}$C/$^{13}$C and $^{16}$O/$^{17}$O ratios.

The isotopic ratios are found from logarithmic shifts. The uncertainty
in the log is of course asymmetric in linear units. In the case of
the carbon isotope ratios most authors replace the asymmetric
uncertainty by a symmetric one.  The absolute differences are
typically small.  Symmetric uncertainty was also adopted for the
oxygen isotopes in this paper for the sake of consistency and easier
readability.  The uncertainty given is the mean between the two
uncertainties, i.e., the measured uncertainty is larger in the
positive direction and smaller in the negative direction than the
listed value.

\subsection{Dependence of results on pulsational phase}

Third dredge up episodes in AGB stars are driven by brief helium shell
flashes.  While the time for these flashes can be of order a year, the 
resulting changes in surface luminosity \citep{wood_zarro_1981} and 
the predicted duration of the associated dredge up are $\sim$10$^2$ times 
longer \citep{pols_et_al_2001}.  Hence we expect the surface abundances
of AGB stars to be constant over time spans of a decade.  Furthermore 
we note that none of the stars discussed in this subsection is undergoing
systematic period change and so are all likely in the interpulse
phase between dredge-up episodes.  We take advantage of this
to test the robustness of our reduction process. 

Atomic and molecular spectra in LPVs are affected in a complex way 
by the large amplitude stellar pulsation \citep{lebzelter_et_al14}.
The pulsation periods of these star are on the order a year.
Changes of velocity and excitation temperature of the CO lines with
phase are well documented.  Near maximum light the lines double as
a shock dissociates the molecular constuents in the atmosphere
\citep{hinkle_et_al_1984}.  We investigated the impact of stellar
variability on our measurements of the isotopic ratios by analyzing multiple spectra
in time series of observations.  We selected four stars, R\,Cas,
T\,Cas, o\,Cet and $\chi$\,Cyg, with extensive time series.  The
1.6~$\mu$m region of these spectra was previously analyzed by
\citet{hinkle_et_al_1982} and \citet{hinkle_et_al_1984}.  In these
two papers measurements of $^{12}$C$^{16}$O and $^{13}$C$^{16}$O
lines were used to determine excitation temperature as a function
of phase.  Here we use these data to investigate the phase dependence
of $^{12}$C/$^{13}$C.

\begin{figure} 
\epsscale{0.7}
\plotone{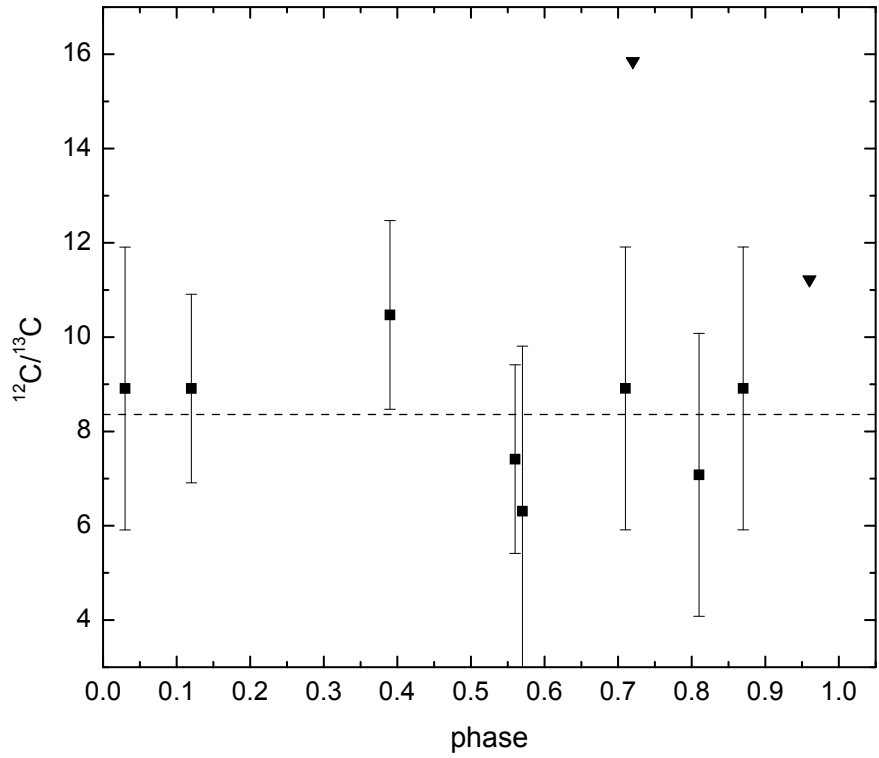}
\caption{
$^{12}$C/$^{13}$C and associated uncertainties plotted against the phase of the measurement
for the M-type Mira R Cas.  For two phases (triangles)
only an upper limit of the ratio could be determined.  Dashed line
is the mean value.\label{f:multiepoch}
}
\end{figure}

Figure \ref{f:multiepoch} shows the $^{12}$C/$^{13}$C ratio versus
phase for the Mira R\,Cas.  R\,Cas has a near-IR spectrum with a
very strong contribution from water \citep{hinkle_et_al_2000}.
Water has a very rich spectrum in the 1.5-2.5~$\mu$m region with
stronger lines concentrated around 1.8 and 2.5~$\mu$.  The presence
of blending water lines makes this star a stringent test since the
selection of CO lines is not as large as in many other Miras.  In 
fact for R Cas we could not find any clean lines for the rare 
oxygen isotopes. For
two phases around maximum light we can derive only an upper limit
on the $^{12}$C/$^{13}$C ratio.  In spite of this any dependence
of the carbon isotopic ratio on phase is weak. We found similar
results for the other stars.  Spectra showing line doubling, usually
appearing around maximum light in Miras, result in smaller line
lists and correspondingly larger uncertainties.

\begin{figure}
\epsscale{0.7}
\plotone{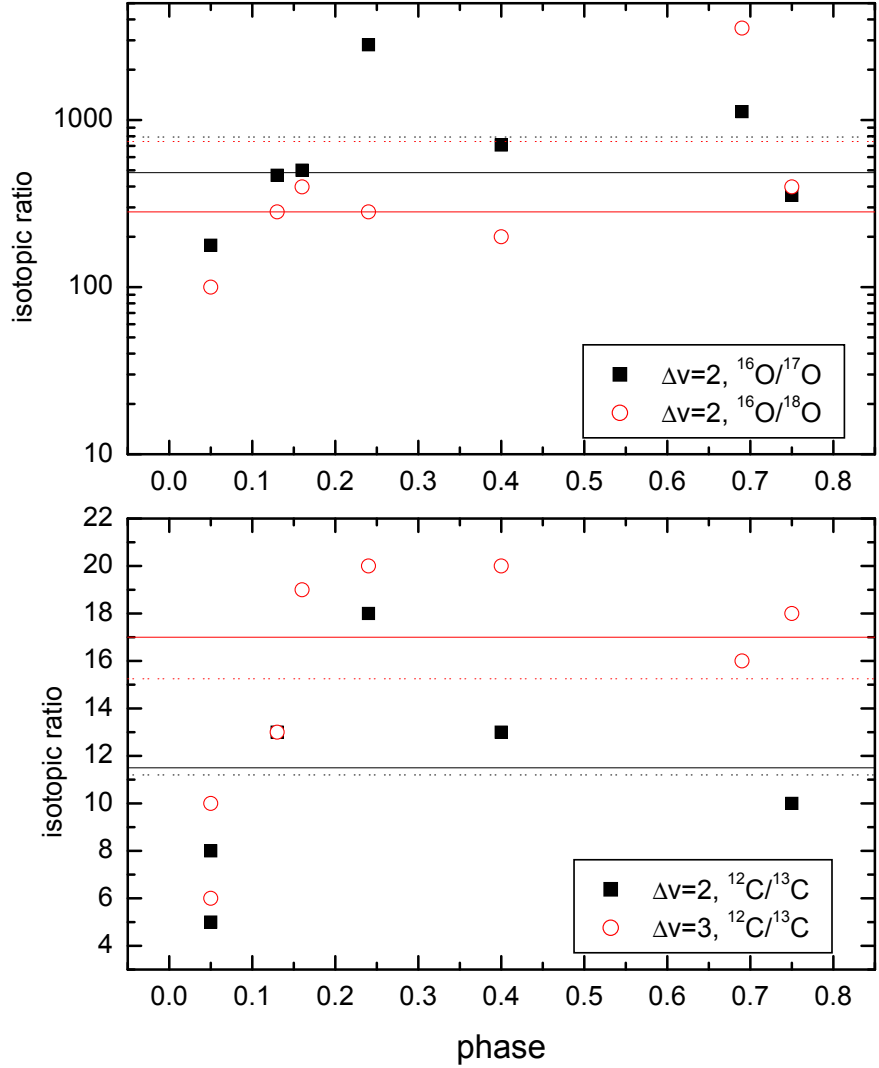}
\caption{Upper panel: Oxygen isotope measurements for o\,Cet as 
a function of phase.  
Heavy solid line is median value for $^{16}$O/$^{17}$O.  Light solid line 
is median value for $^{16}$O/$^{18}$O.  Dotted lines are the means.
Lower panel:
Carbon isotopic ratios from the first and second overtone lines as 
as function of phase.  Solid and dotted lines are medians and means.
\label{f:ocet_iso}
}
\end{figure}

The archival measurements, however, cover only the $H$-band. 
Phase dependence of the oxygen isotopes, where the lines are in the 
$K$-band, could not be studied with these data.  To investigate the
oxygen isotopes we measured a set of spectra of o\,Cet.  o\,Cet has
a spectrum that perhaps is a more difficult observational case 
for determining $^{12}$C/$^{13}$C than R Cas.
The second overtone $^{13}$C$^{16}$O lines are fairly weak and,
similar to  R\,Cas, the CO spectrum is blended with a large number of
water vapor lines.  In Figure \ref{f:ocet_iso} the carbon and oxygen
isotopes are shown as a function of phase. Clearly phases near
maximum light give divergent values.  The stronger first overtone
$^{13}$C$^{16}$O lines are also less reliable near maximum light
due to line doubling.  Overall the oxygen isotopic ratios have more
scatter between measurements than is the case for the carbon isotope.
This is not surprising since the line list for the oxygen isotopes
is at least an order of magnitude smaller than the line list for
the second overtone $^{13}$CO lines.

For most stars with multiple observations in our sample the deviation
from the mean value does not exceed $\pm$38\,\%. For a few stars
like U Ori or X Oph the difference is larger. For Table \ref{t:results}
we used the mean value in case of multiple observations.  Note that
we combine measurements from various pulsation cycles in our analysis,
so that variability effects can be the result of either inter-
or intra-cycle variations.

\subsection{Uncertainties}\label{symerr}

The uncertainties given in Table\,\ref{t:results} result primarily
from the width of the CoG due to the scatter in equivalent widths.
In the case of multiple observations, the differences between the
results for each observation are also included in the uncertainty.
The scatter in the CoG equivalent widths can be attributed to noise,
unidentified line blends, and continuum placement.  Continuum
placement contributes scatter since the lines span a significant
wavelength range with lines of different strengths intermingled.
The equivalent widths are determined from a fit to the depths versus
width relation which minimizes the effect of blending in the line
widths.  For the $^{13}$C isotope, with a rich selection of lines,
blends are likely a minor problem in the analysis.  Of course, the
opposite is true for the $^{18}$O isotope where the result depends
on only a few lines or even a single line.  This effect is very
difficult to quantify and may lead to significantly different
isotopic ratios in individual stars.  Another possible source of
uncertainty is the excitation temperature used.  Tests varying 
the excitation temperature by $\pm$200\,K result in changes in the
isotopic ratio of about 10\,\%. Larger deviations are unlikely since
this typically leads to obvious distortions in the CoG.  Comparisons
of these results and results obtained from spectra at different
phases and consequently different excitation temperatures indicate
that the uncertainties given in Table\,\ref{t:results} are reasonable
statistical estimates.

\subsection{Comparison with literature data}\label{s:comp lit}

Isotopic ratios were previously determined for a subset of our
sample. In Table \ref{t:complit} we compare these literature values
with our findings.  In a few cases more than one $^{12}$C/$^{13}$C
value can be found in the literature.  Agreement between literature
values is occasionally poor, especially for three of the carbon
stars, R\,Lep UX\,Dra, and V\,Hya.  For the few cases where we found
more than one oxygen isotopic ratio in the literature, we again
found a range of values.

\begin{figure}
\epsscale{1.0}
\plotone{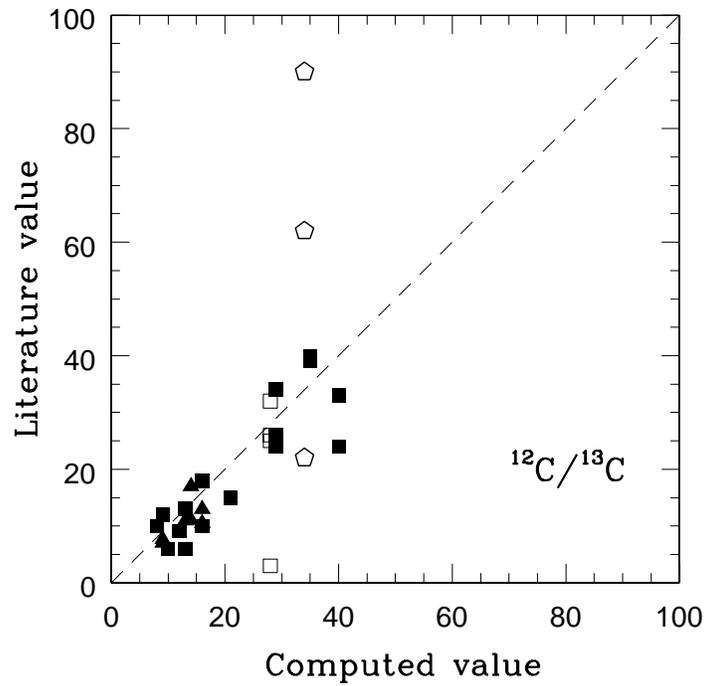}
\caption{
$^{12}$C/$^{13}$C values computed in this paper are compared to literature values.
Non-variable stars are filled triangles.  The two stars with large disagreements in
the literature, UX Dra and R Lep, are shown with open squares and open pentagons, respectively.
\label{f:comparison}
}
\end{figure}

Our carbon isotopic ratios are generally in reasonable agreement
with the literature values (Figure \ref{f:comparison}).  A notable
exception is V\,Hya where we derived ratio of 200$\pm$50.  This is
much larger than the value of 69 found by \citet{lambert_et_al86},
71 found by \citet{milam_et_al_2009}, and 33 found by
\citet{jura_et_al_1988}.  Agreement with the literature is clearly
worse for the oxygen isotopic ratios. However, many fewer values
appear in the literature.  While there are a few stars where we
derived very different values several measurements agree within the
uncertainties. An example is the star W\,Hya, where \citet{khouri_et_al14b}
derive almost identical values to our values using completely
different lines.

As a final test we carried out the entire analysis on four M giants
that were not LPVs.  The stars selected have established values for
the carbon and oxygen isotopes and were observed with the same
experimental setup as the LPVs.  The observations are listed at the
bottom of Table \ref{t:obslog}.  The results are given at the bottom
of Table \ref{t:complit}.  For the M giants there is excellent
agreement with the literature.

\section{STELLAR EVOLUTION AND NUCLEOSYNTHESIS OF C AND O ISOTOPES}

From the evolution of surface abundances, AGB stars can be roughly
divided into three groups\footnote{ The mass bounds of the three
groups depend on many parameters such as initial
composition and efficiency of mass loss both before and
during the AGB phase. The values discussed here, representative of
solar metallicity stars, are presented for guidance purposes.}:

\begin{itemize} 

\item Main sequence mass \textless~$1.5$~$M_\odot$. These stars do 
not retain sufficient envelope mass to experience TDU
events during the AGB phase. Their AGB surface composition is largely
determined by the first dredge-up and, eventually, by the RGB
extra-mixing.

\item Main sequence mass
between $1.5~M_\odot$ and $3~M_\odot$. These stars undergo repeated strong TDU episodes that bring
to the surface the products of the He burning. The number of TDU
episodes and the amount of material dredged-up depend on the
efficiency of mass loss before and during the AGB phase. They are among the most 
important nucleosynthesis sites in all galaxies.

\item Main sequence mass $> 3$ $M_\odot$. These stars undergo a second
dredge-up, shallow TDU episodes, and, possibly, HBB.

\end{itemize}

The `super-AGB stars' belong to the last group with
main sequence masses between $8~M_\odot$ and $10~M_\odot$. After the core He burning these stars
ignite carbon in a highly degenerate core and at the end of the
carbon burning experience an AGB phase whose characteristics are
similar to, but more extreme than, those of the AGBs with $M\sim 6-7~M_\odot$. In
particular super-AGB stars are expected to experience
deep HBB with temperature at the bottom of the convective envelope
exceeding $10^8$ K \citep[see][and references therein]{doherty2010}.

In this section we summarize the results of extant nucleosynthesis
calculations.  Differences exist among the various models available
in the literature \citep[e.g.,][]{lugaro2003}.  These stem from
different choices for initial composition, numerical recipes, and
values for parameters such as nuclear reaction rates \citep[for more
details see section 2 in][]{straniero2014}.  Despite these
differences the overall scenario is quite well established.  In
the following we report numerical values of the various isotopic
ratios as derived from the FRUITY database
\citep[][http://fruity.oa-teramo.inaf.it/]{cristallo2015}.  In
general we will discuss results of models whose initial composition
is solar.

\subsection{Carbon isotopes}

The CNO cycle depletes the most abundant carbon isotope, $^{12}$C, and
enhances $^{13}$C.  In the opposite sense, He burning is the main producer
of $^{12}$C. During stellar evolution the ashes of these
fundamental nuclear processes are mixed with the pristine material
in the stellar envelope.  Mixing processes of this type, also called
dredge-up episodes, are well known phenomena taking place in giant
stars. Extra-mixing processes induce further modifications of the
surface composition.

\begin{figure} 
\epsscale{0.7}
\plotone{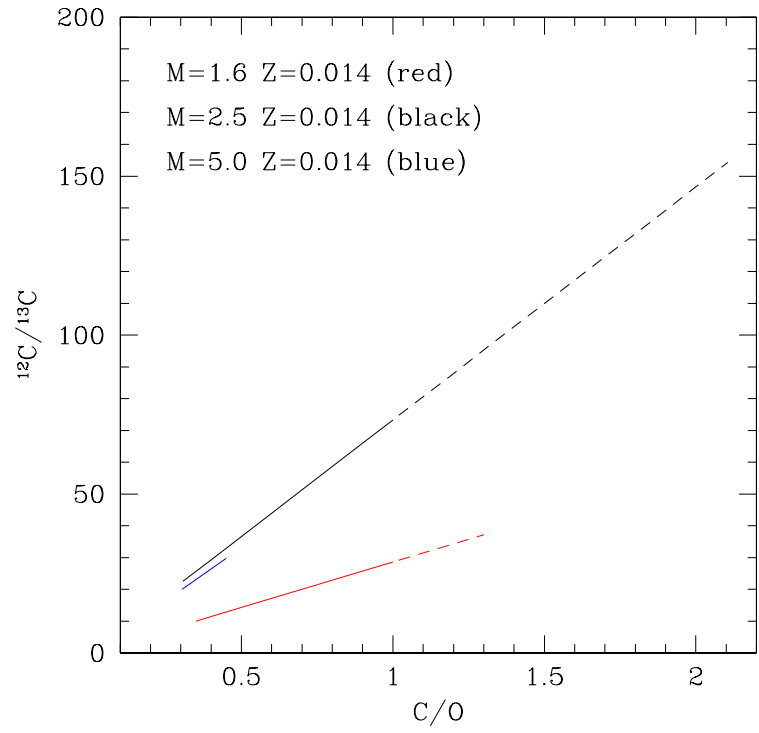}
\caption{$^{12}$C/$^{13}$C versus C/O for three AGB stellar models
of different mass.\label{fig1_teo}}
\end{figure}

The carbon isotopic ratio of the sun is about 90 with galactic main
sequence stars expected to share similar values.  As a star leaves
the MS the first dredge-up causes a reduction of the carbon isotopic
ratio down to $\sim$20 $-$ 25.  In stars with mass \textless$2$ M$_\odot$ further
reduction of the $^{12}$C/$^{13}$C by up to a factor of two can
occur as the result of RGB extra-mixing.  The origin of RGB extra-mixing
is unclear but its existence is confirmed by observational evidence
\citep[for a review, see][]{charbon1999}.  During the early-AGB
phase stars with mass $\ge4$ M$_\odot$ experience a second dredge-up.
Shell H-burning temporarily ceases and external convection penetrates
into the H-exhausted core. In this case, both carbon isotopes are slightly
depleted so that their ratio remains almost unaltered. Therefore,
$^{12}$C/$^{13}$C ratios between 10 and 25 are expected at the onset
of the first thermal pulse in AGB stars with different masses.
The lower values, 10~\textless~$^{12}$C/$^{13}$C~\textless~20, are expected for
stars with main sequence mass \textless2 $M_\odot$.  The slightly higher values, 
20~\textless~$^{12}$C/$^{13}$C~\textless~25,
are expected for the more massive stars.  For stars undergoing recurring third
dredge-up episodes the carbon isotopic ratio increases as fresh $^{12}$C
synthesized in the He-burning shell is mixed to the surface. The
continued addition of carbon to the stellar surface layers can turn an
M-type into a C-type star.

Although all AGB stars are expected to experience thermal pulses
only a few are followed by third dredge-up episodes.  As discussed
in \citet{straniero2003} the occurrence of TDU and its efficiency
in bringing the ashes of the internal nucleosynthesis to the stellar
surface depends on the core mass ($M_H$) and the envelope mass
($M_E$).  The third dredge-up is inefficient in AGB stars with
$M_E$~\textless~0.5 or $M_H~>~0.9$~M$_\odot$.  The first condition implies the
existence of a minimum stellar mass for the occurrence of the TDU.
The precise value of this threshold depends on the mass lost before
the beginning of the thermal pulse phase.  For solar metallicity
the limiting value of the main sequence mass should be $\sim$ $1.3-1.4$
$M_\odot$.  Marginal TDU episodes are also seen in models of 
massive AGB stars that develop quite large core masses \citep[this
issue has been extensively discussed in][]{straniero2014}.  As a
consequence AGB stars whose initial mass is $ \lesssim~1.5~M_\odot$
and $ \gtrsim~5~M_\odot$ should appear as M stars\footnote
{A more rapid increase of the C/O ratio is expected at low metallicity, 
even in AGB stars undergoing a few shallow TDU episodes. In the more massive AGBs, however,
the HBB prevents the development of high C/O ratios.}. 
Rather deep
dredge-up episodes are experienced by stars with masses between 1.5
and 3 $M_\odot$ so these stars may evolve into S and, possibly, C
stars.  The mass-loss rate determines the AGB lifetime and, hence,
the maximum number of TDU episodes plays a fundamental role in the
evolution to M-S-C spectral type.

The maximum $^{12}$C/$^{13}$C attained by this class of AGB stars
also depends on the composition at the beginning of the AGB.  Figure
\ref{fig1_teo} illustrates three example AGB models of initial mass
1.6, 2.5, and 5 $M_\odot$.  O-rich and C-rich phases are plotted
as solid and dashed lines, respectively.  In the 1.6 $M_\odot$ model
(red line) the star arrives on the AGB with $^{12}$C/$^{13}$C$=10$
and C/O$=0.35$.  These values are determined by the first dredge-up
followed by RGB extra-mixing.  No RGB extra-mixing has been considered
in the 2.5 $M_\odot$ and 5 $M_\odot$  models. For these models
$^{12}$C/$^{13}$C is $\sim21$ and C/O$\sim0.31$ at the beginning of the
AGB.  Both ratios then increase because of the TDU. The final values
attained at the AGB tip are $^{12}$C/$^{13}$C$=37$ and C/O$=1.3$
for M=1.6 $M_\odot$, $^{12}$C/$^{13}$C$=154$ and C/O$=2.1$ for M=2.5
$M_\odot$, and $^{12}$C/$^{13}$C$=30$ and C/O$=0.45$ for M=5
$M_\odot$.  The two less massive models become carbon stars,
i.e., attain C/O$=1$, when $^{12}$C/$^{13}$C is 28 (1.6 $M_\odot$) and
71 (2.5 $M_\odot$).

In practice intrinsic C-stars, i.e., those C-stars whose origin is
a direct consequence of the ongoing third dredge-up, should have
carbon isotopic ratios $>25$. Lower value carbon isotopic ratios
in C-stars have at least two explanations.  One scenario is that
an extrinsic process, for instance mass exchange or merger, caused
C/O$>1$.  
J-type carbon stars are likely extrinsic C-stars
\citep{abia_isern_1997, abia_et_al_2003}.
The second scenario is that the low $^{12}$C/$^{13}$C ratio is a consequence of an extra-mixing
process taking place during the AGB \citep[see][]{abia_et_al01, lebzelter_et_al08}.
Intrinsic
S-stars, where C/O is very near but perhaps not quite 1, should
have $^{12}$C/$^{13}$C~\textless~70.   O-rich stars with initial mass 
\textless~$1.5~M_\odot$ should have $^{12}$C/$^{13}$C~$\sim~10-15$, as determined
by the combined effects of the first dredge-up and the RGB extra-mixing.
Slightly larger values, i.e., $^{12}$C/$^{13}$C$~\sim~25-35$, are
found for masses between $4~M_\odot$ and $6~M_\odot$. However, if the initial mass
is $>6~M_\odot$, then the temperature at the base of the convective
envelope may become large enough to activate the HBB.  As
a consequence, the $^{12}$C/$^{13}$C ratio decreases.  Extreme cases
are found in models of super-AGB stars ($M~=~8 - 10$ $M_\odot$)
undergoing a very efficient HBB.  They show $^{12}$C/$^{13}$C as
low as  $3 - 5$, the typical equilibrium value of the CN cycle
\citep[e.g.,][]{doherty2014}.

\subsection{Oxygen isotopes}

\begin{figure} 
\epsscale{0.7}
\plotone{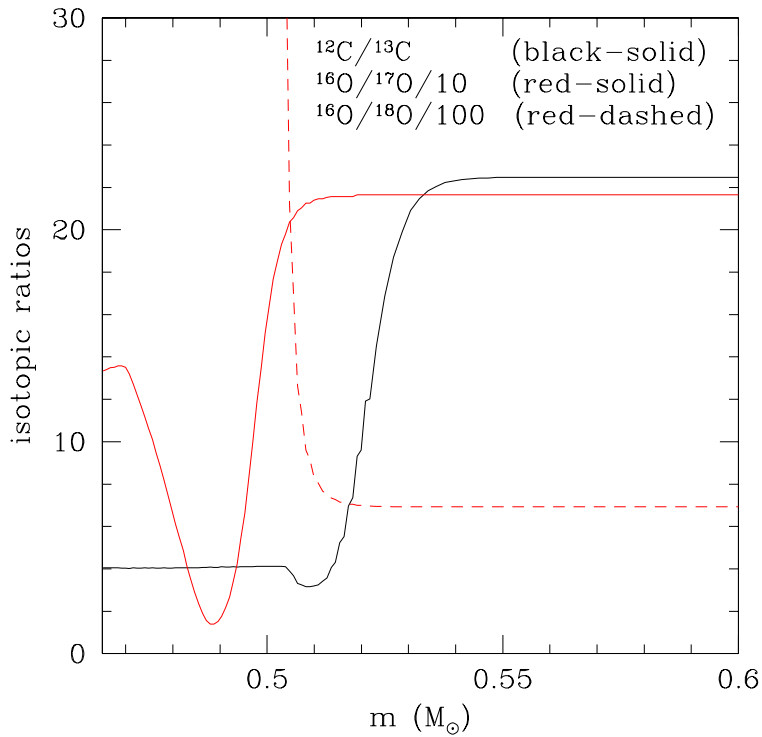} 
\caption{Carbon and oxygen isotopic ratios in the innermost layer of the
H-rich envelope of a 2 $M_\odot$ RGB model. The variation of the
$^{12}$C/$^{13}$C starts in a relatively external region where the
temperature is $T>10$ MK.  The variations of the oxygen isotopes take
place more internally where $T>15$ MK.
\label{profile_2p0}} 
\end{figure}

$^{18}$O is depleted by H burning whereas $^{17}$O is enhanced.
$^{16}$O is depleted but at a rather high temperature.  On the
contrary, both $^{16}$O and $^{18}$O are produced by He burning
through the $^{12}$C$(\alpha,\gamma)^{16}$O and 
$^{14}$N$(\alpha,\gamma)^{18}$F$(\beta)^{18}$O reactions.  Solar
values of the oxygen isotopic ratios are $^{16}$O/$^{17}$O$=2700$
and $^{16}$O/$^{18}$O$=490$ \citep{lodders_et_al09}.  
In a previous paper \citep{lebzelter_et_al15}
we discussed theoretical predictions and uncertainties for
the oxygen isotopic ratios expected after the occurrence of first
dredge-up.  Both oxygen isotopic ratios are sensitive to the initial
composition.  At variance with the $^{12}$C/$^{13}$C, the RGB
extra-mixing has negligible effects on the oxygen isotopic ratios
\citep[see][]{abia_2012}. The reason is illustrated in Figure
\ref{profile_2p0}, in which we plot the variation of the C and 
O isotopic ratios in a typical H-burning region, i.e., the innermost
layer of the H-rich envelope of a 2 $M_\odot$ RGB model. From
this plot it can be seen that extra-mixing deep enough
to modify the abundances of the oxygen isotopes would imply
$^{12}$C/$^{13}$C~\textless~5. Note that where the $^{12}$C/$^{13}$C is
reduced to $\sim 10$ the oxygen isotopic ratios are not changed by the
CNO cycle.

Assuming a solar composition the variation of oxygen isotopes with
mass after the first dredge-up can be summarize as follows.
$^{16}$O/$^{17}$O is reduced down to a minimum value of $\sim$200
in stars with $M\sim 2.5$ $M_\odot$.  For smaller stellar masses
the $^{16}$O/$^{17}$O is larger.  For instance we find
$^{16}$O/$^{17}$O$~=~1000$ and 2650 for models of 1.5 $M_\odot$ and 1
$M_\odot$, respectively. For $M>2.5 M_\odot$ the $^{16}$O/$^{17}$O ratio
slightly increases with mass.  We find $^{16}$O/$^{17}$O$~=~340$ and
430 for models of 4 and 6 $M_\odot$, respectively.  The variation
of $^{18}$O with mass is different from $^{17}$O with $^{16}$O/$^{18}$O
increasing after the first dredge-up because of $^{18}$O consumption
taking place in the H-burning zone.  Values of about 700 are found
for $M\ge2$  $M_\odot$ and between 500 and 700 for smaller stellar
masses.

The second dredge-up produces smaller modifications of the oxygen
isotopes than the first dredge-up.  In a 6 $M_\odot$ model, we found
405 and 720 for $^{16}$O/$^{17}$O and $^{16}$O/$^{18}$O, respectively,
with 430 and 700 after the first dredge-up.  Similarly, the third
dredge-up does not substantially modify the abundances of the oxygen
isotopes \citep{cristallo2015} unless a substantial overshoot at
the base of the convective zone, powered by a thermal pulse, allows
deep mixing of primary $^{16}$O.  If this is the case then both
the oxygen isotopic ratios should increase at the stellar surface.
However, the consequences of this overshoot on the bulk of the s-process
nucleosynthesis would be dramatic \citep{lugaro2003}. Indeed, even in low mass stars, the
maximum temperature attained at the bottom of the convective zone
would be large enough to activate the $^{22}$Ne($\alpha$,n)$^{25}$Mg
reaction. The resulting high neutron density generated by this
neutron source would imply a substantial modification of the s-process
path at the main branching points. Large amounts of Rb and rare
isotopes, like $^{96}$Zr, would be synthesized which is not supported
by abundance analysis of AGB stars and presolar grains
\citep{lambert_et_al_1995, abia_et_al01}.

HBB might affect the oxygen isotopes in the more massive
AGB stars.  The most important effect is a significant depletion
of the $^{18}$O. As a result $^{16}$O/$^{18}$O could became as
large as $10^6$ \citep{boothroyd_1995}. In the most extreme cases
$^{16}$O and $^{17}$O could also decrease. It should be recalled,
however, that a quantitative analysis of HBB is still
hampered by many uncertainties.  Briefly, the nuclear burning
timescales of the various isotopes involved in the nucleosynthesis
processes vary over several orders of magnitude. Some of these
timescales are smaller, some are comparable to, and some 
definitely larger than the mixing timescale, i.e., the time required
to move a nucleus from the bottom to the top of the convective
envelope. As a consequence the results of the nucleosynthesis
calculations are sensitive to the details of the adopted mixing
algorithm, in particular to those related to the coupling of mixing
and burning.

Summarizing, with the exception of the more massive AGB and super-AGB
stars undergoing HBB the oxygen isotopic ratios in LPVs should
coincide with those left by the first dredge-up.  No significant
differences are expected among M, S, and C stars having similar mass.
The majority of the intrinsic S and C stars, typically having masses 
ranging between 1.5 and 3 $M_\odot$, should have $^{16}$O/$^{17}$O
roughly confined between 200 and 1000.  M stars, which can be less
massive, can show larger $^{16}$O/$^{17}$O ratios up to about 2700 
\citep[see also models by][]{karakas_et_al_2010}.
On top of the variations of these ratios with
stellar mass a spread of the observed values is likely due to
a spread of the initial abundances of the oxygen isotopes.

\begin{figure} 
\epsscale{0.7}
\plotone{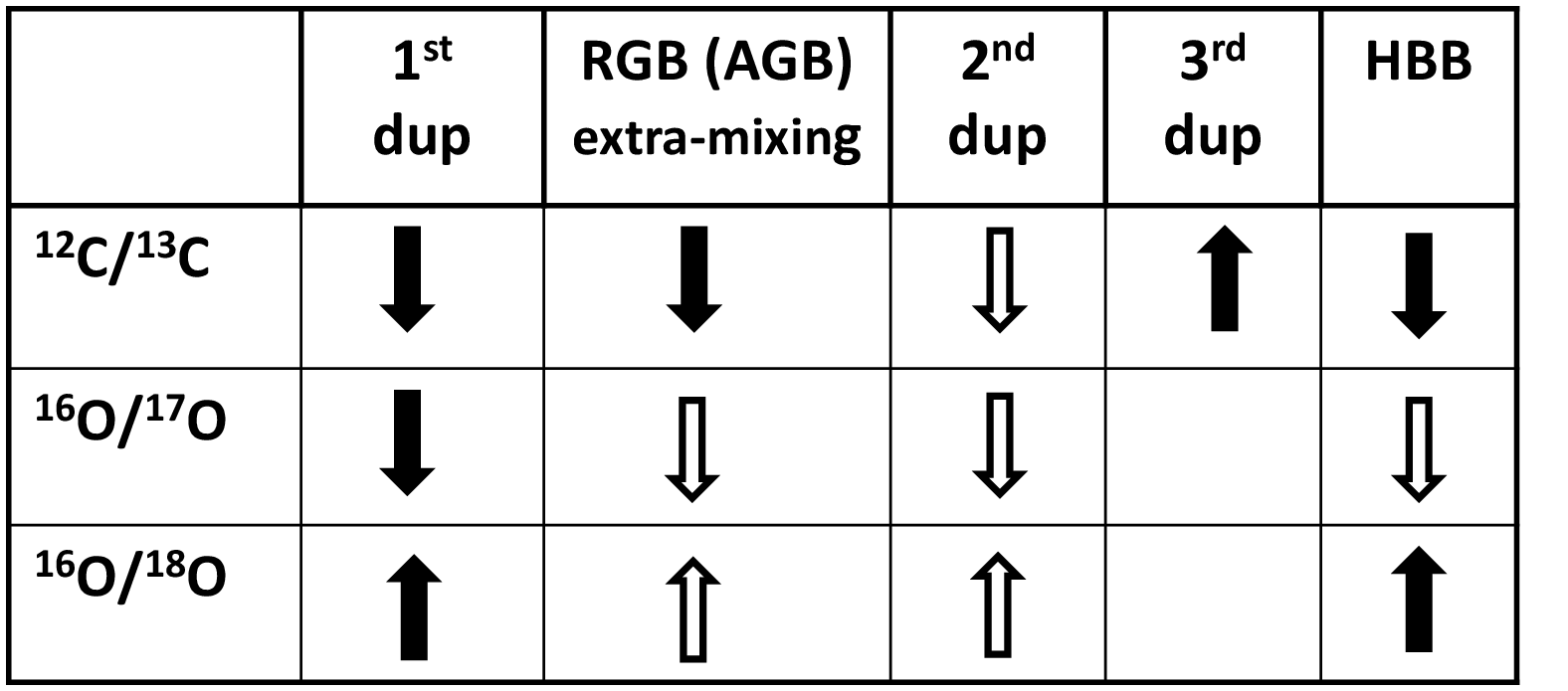} 
\caption{Influence of the various mixing events on the carbon and
oxygen isotopic ratios. Filled arrows indicate a significant variation
of the isotopic ratio while open arrows indicate small variations (smaller
than the observational errors).
\label{tab_mix}} 
\end{figure}

The theoretical scenario described so far is summarized in Figure
\ref{tab_mix}.  As a consequence of a deep mixing process the
$^{12}$C/$^{13}$C usually decreases except in the case of the third
dredge-up. $^{16}$O/$^{17}$O can only decrease but $^{16}$O/$^{18}$O
always increases. Therefore $^{16}$O/$^{18}$O ratios smaller than
the solar value, like those observed in open cluster giants
\citep{lebzelter_et_al15}, necessarily indicate a variation of the
oxygen isotopic composition with respect to the solar, in particular
$^{18}$O enhanced and/or $^{16}$O depleted.  Moreover, the
$^{16}$O/$^{17}$O ratio is a good tracer of the stellar mass.

\section{DISCUSSION}

\subsection{Carbon isotopes}

\begin{figure} 
\epsscale{0.8}
\plotone{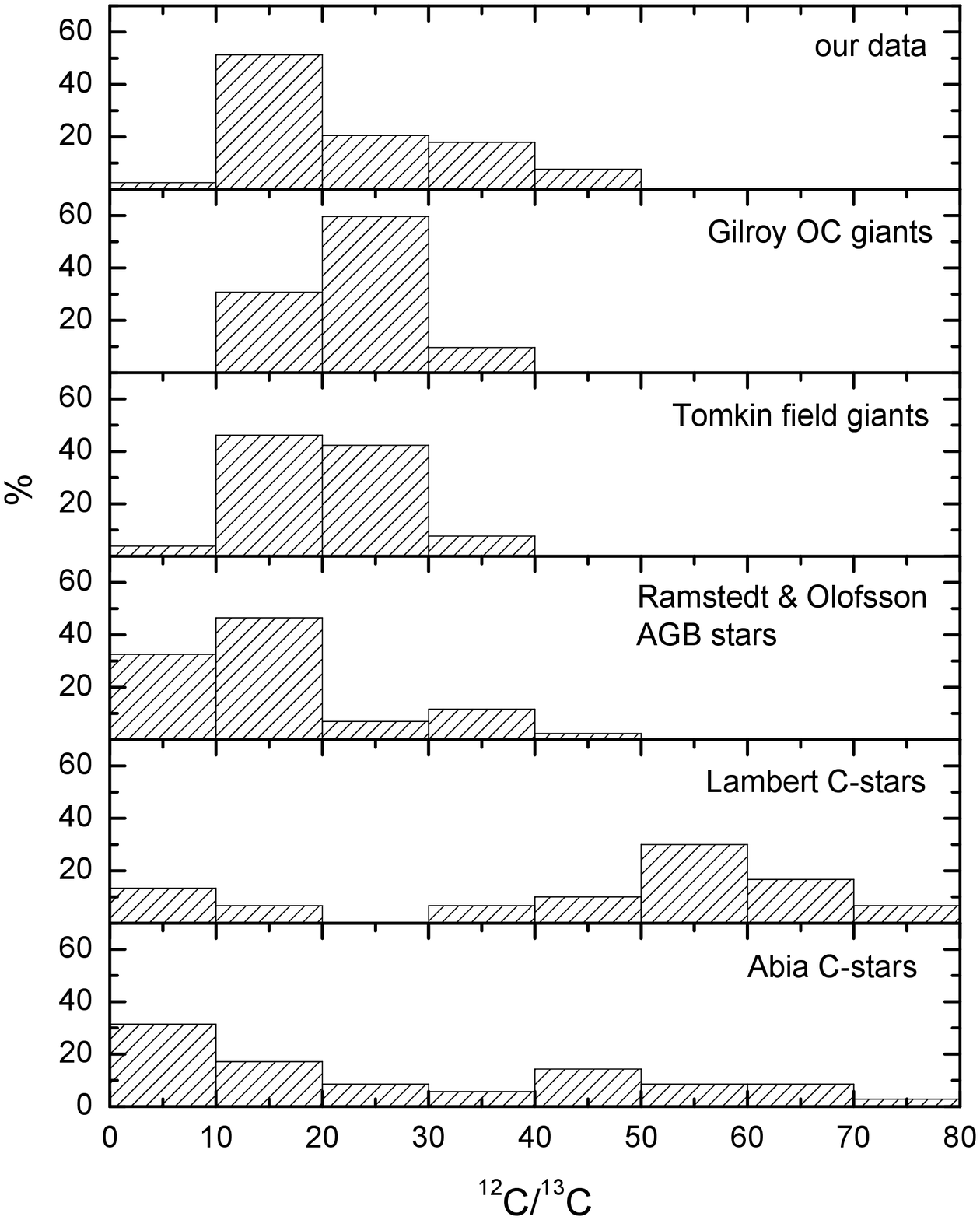} 
\caption{Histograms of
the $^{12}$C/$^{13}$C distribution in various samples of evolved
stars. From top to bottom: (1) Data presented in this paper. (2)
\citet{gilroy_89}. (3) \citet{tomkin_et_al76}.  (4)
\citet{ramstedt_olofsson14}. (5) \citet{lambert_et_al86}. (6)
\citet{abia_isern00} and \citet{abia_et_al01}. See text for details.
\label{f:cisopercent}} 
\end{figure}

The $^{12}$C/$^{13}$C ratio is often used as an indicator of mixing
processes in evolved stars.  It is measurable using a number of
different observational approaches.  $^{12}$C/$^{13}$C is typically
smaller than the oxygen isotopic ratios and hence more easily
determined.  Large sets of values, primarily for non- or only
slightly variable evolved stars, are available in the literature.
In Figure \ref{f:cisopercent} we compare the distribution of the
$^{12}$C/$^{13}$C for our sample with $^{12}$C/$^{13}$C distributions
for giants, AGB stars and carbon stars in published data sets.

The samples compared in Figure \ref{f:cisopercent} cover very
different groups of evolved stars. Our data are dominated by large
amplitude LPVs with most stars of M-type.  Four of our sample stars are classified
as spectral type S, six as spectral type C, and just one as type SC. The
Mira S\,Cep has a $^{12}$C/$^{13}$C ratio too large to be shown on
this plot.  The peak in our distribution (top panel of
Figure\,\ref{f:cisopercent}) is for $^{12}$C/$^{13}$C between 10 and
20.  This indicates a sample biased toward low-mass stars.
This is in contrast to the second sample plotted in
Figure\,\ref{f:cisopercent} which is from \citet{gilroy_89} and
consists of late-type giants in open clusters. The \citet{gilroy_89} distribution
peaks between 20 and 30.  Indeed, the stars in the \citet{gilroy_89}
sample are progenitors of AGB stars with masses between 1.6 and
6\,M$_{\sun}$.  Another sample of likely AGB progenitors is the one
by \citet{tomkin_et_al76} consisting of late-type field giants.
This sample likely includes a mixture of low- and intermediate-mass
stars. In agreement with this assumption the $^{12}$C/$^{13}$C
values concentrate between 10 and 30.

The sample of \citet{ramstedt_olofsson14} is dominated by
AGB stars. Indeed there are several stars in common between the
\citet{ramstedt_olofsson14} sample and our sample (see Table
\ref{t:complit}).  However, unlike our sample the 
\citet{ramstedt_olofsson14} sample
has M-, S-, and C-type stars roughly
equally represented. The histogram (fourth panel from the
top in Figure\,\ref{f:cisopercent}) peaks between 10 and 20 like our
data. Interestingly, a larger fraction of stars with $^{12}$C/$^{13}$C\textless10
are present in the  \citet{ramstedt_olofsson14} sample than in our sample.
At the same time, the fraction of stars with $^{12}$C/$^{13}$C$>$20
is lower. It is worth noting that the study of \citet{ramstedt_olofsson14}
is the only one in our comparison that is based on radio data while
all the other authors used optical and/or infrared spectra for their
analysis. The use of radio data implies that the stars are  
undergoing significant mass loss.

The final two literature samples shown in Figure\,\ref{f:cisopercent}
consist of C stars only. Both samples show similar
histograms with two peaks, one at very low $^{12}$C/$^{13}$C ratios
and one above 40.  
There are a few
C-stars with $^{12}$C/$^{13}$C$>$80 which are
not plotted for the sake of clarity.
The sample of \citet{abia_isern00} consists of J-type carbon stars
with $^{12}$C/$^{13}$C, by definition, lower than 15.  Nevertheless, in 
the sample of \citet{abia_et_al01} there are 
C stars classified as N-type with $^{12}$C/$^{13}$C$<$25. Such an occurrence 
is considered a proof of the occurrence of AGB extramixing.

\subsection{Oxygen isotopes}

\begin{figure} 
\epsscale{0.8}
\plotone{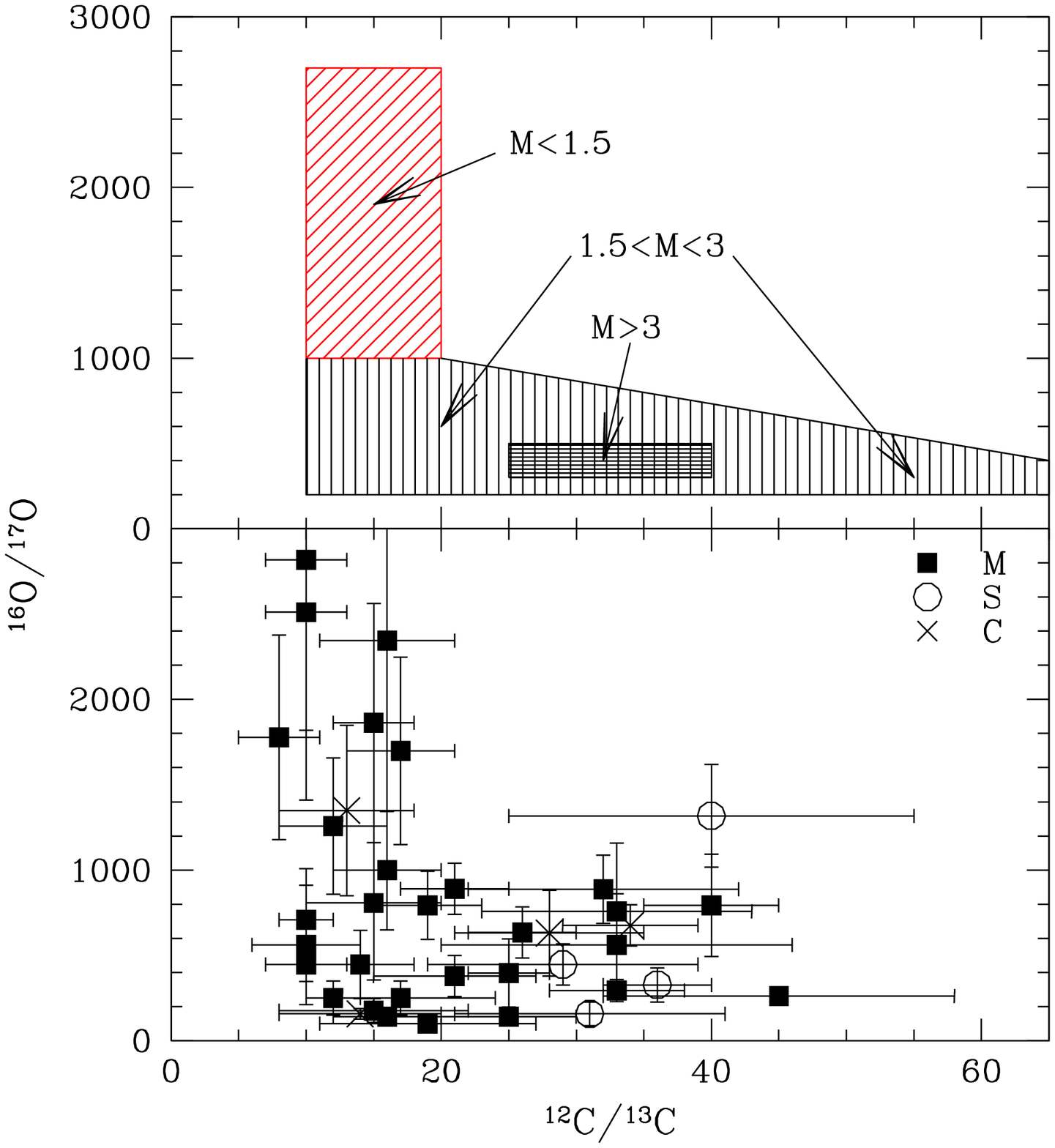} 
\caption{
$^{16}$O/$^{17}$O versus $^{12}$C/$^{13}$C diagram. Upper panel:
Shaded areas delimit the parameter space where we expect to find
AGB stars with the three labeled mass ranges. The location of AGB
stars with M$>$6 is not shown.  For such stars, the occurrence of
the HBB should imply quite low values of both $^{16}$O/$^{17}$O and
$^{12}$C/$^{13}$C. Lower panel:  Derived values for the sample stars
plotted in the parameter space seen in the upper panel.  Note the
different symbols used for M-type, S-type, and C-type stars.
\label{o17vc13}} 
\end{figure}

In addition to $^{12}$C/$^{13}$C, $^{16}$O/$^{17}$O is also a good
stellar mass tracer.  The upper panel of Figure \ref{o17vc13} shows
the areas in the $^{12}$C/$^{13}$C versus $^{16}$O/$^{17}$O diagram
occupied by AGB stars of different masses.  Values for stars included
in our sample are shown in the lower panel.  Our sample stars cover
the same range in this parameter space expected for AGB stars with
masses below 3 M$_{\sun}$.  The selection of our sample was driven
primarily by apparent near-infrared brightness 
and variability.  The sample reflects a
mass range typical for Miras in the solar neighborhood.  The number
of objects with $^{16}$O/$^{17}$O below 1000 is clearly larger than those
with higher $^{16}$O/$^{17}$O indicating that the majority of the
Miras have main sequence masses $\sim 1.5$ M$_{\sun}$.  
The typical mass range is then around 1.5
to 2\,M$_{\sun}$ for most Miras with a smaller fraction having
masses of less than 1.5\,M$_{\sun}$. 

\begin{figure} 
\epsscale{0.8}
\plotone{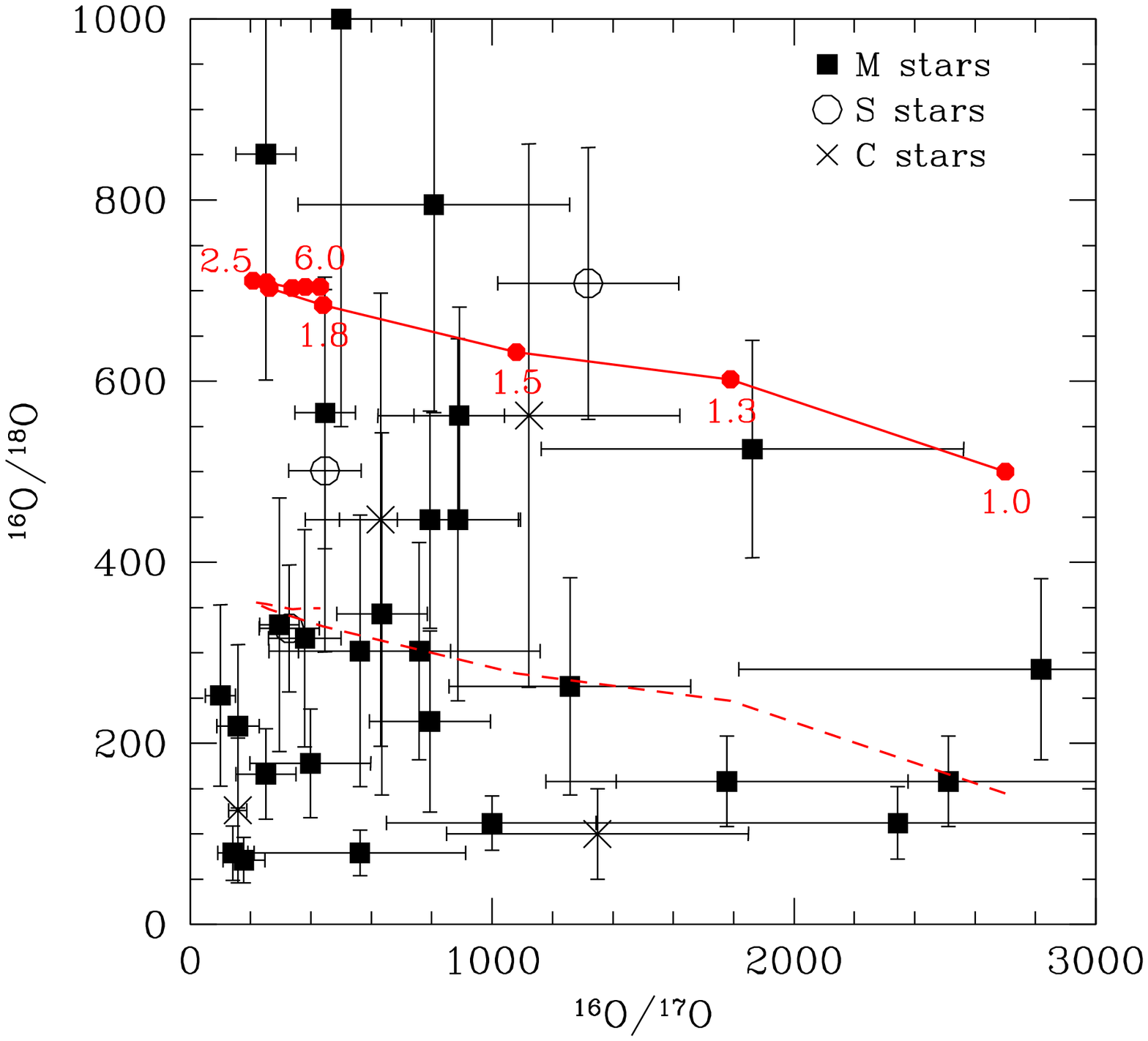} 
\caption{
Positions of the stars included in our sample in the $^{16}$O/$^{18}$O
versus $^{16}$O/$^{17}$O diagram.  The solid line represents model
predictions for solar composition AGB stars of different (indicated)
initial mass.  The dashed line shows this relation for an initial
$^{18}$O abundance twice solar. As in Figure \ref{o17vc13} 
different symbols are used to mark 
the M-,S-, and C-type stars. 
\label{O17vO18}} 
\end{figure}

$^{16}$O/$^{18}$O, on the other hand, is almost insensitive to
initial mass but can be a good tracer of the oxygen composition
either in the gas from which the stars were formed and/or from
particularly deep mixing processes like the HBB. The distribution
of our sample stars in the $^{16}$O/$^{18}$O versus $^{16}$O/$^{17}$O
diagram is shown in Figure \ref{O17vO18}. Model predictions for AGB
stars with solar composition and enhanced original $^{18}$O abundance
are shown. While we have to keep in mind that our $^{16}$O/$^{18}$O
measurements have higher uncertainties than the $^{16}$O/$^{17}$O
and $^{12}$C/$^{13}$C measurements, the diagram suggests that there exists a range
of initial $^{16}$O/$^{18}$O values in the sample stars. 
Either a reduced initial $^{16}$O
abundance or an enhanced initial $^{18}$O abundance compared to the
solar value is required for the majority of the stars. M-, S-, and C-type stars
do not differ in this respect.  The predicted dependency on mass
may be weakly visible in our observations but the measurement
uncertainties are simply too large to reach a clear conclusion.

\subsection{M stars} 

Spectral type M AGB stars are not polluted with third dredge-up
material.  Typically these stars have a mass too low to experience
third dredge-up events.  However, it is possible that the M AGB
stars include stars of higher mass at the beginning of AGB evolution,
i.e., before experiencing a significant change of surface composition
by third dredge-up.  The spectral type alone is insufficient to
further constrain the mass of a particular LPV.
The isotopic ratios permit additional conclusions.  Once the star
has experienced several third dredge-up events, changes in the
abundance of s-process elements permit constraints on mass.  In the
case of a massive AGB star experiencing HBB the C/O and the $^{12}$C/$^{13}$C
can be maintained at low levels in spite of the TDU but some s-process
enhancement will occur.  It is possible that the HBB prevents or,
at least, makes the TDU very weak \citep{straniero2014}.  These objects will be very rare
compared to the less massive AGB.

The majority of the M-type stars in our sample have $^{12}$C/$^{13}$C
between 10 and 20.  They have not experience third dredge-up but
most probably have experienced some extra-mixing on the RGB.  Their
mass should be smaller than $\sim 2$\,$M_\odot$. A part of this
sample shows $^{16}$O/$^{17}$O$>1000$ limiting their mass further
to below 1.5\,$M_\odot$ (see Figure \ref{o17vc13}). About two thirds
of the M-type stars with $^{12}$C/$^{13}$C~\textless~20 show $^{16}$O/$^{17}$O
values more compatible with a mass above 1.5\,$M_\odot$ (lower left
corner of Figure \ref{o17vc13}).  The masses of this set can thus
be constrained quite well to a range between 1.5 and 2\,$M_\odot$.
This is consistent with other mass estimates for Miras in the same period
range recently summarized by \citet{feast_09}.

Our sample is entirely made up of Mira and SRa variables.  These
are fundamental mode AGB pulsators associated with the final stage
of the AGB evolution \citep[see][and references therein]{wood_2015}.
This suggests that our sample stars are not at the beginning of the
AGB phase.  We expect a slow increase of the surface carbon isotopic
ratio as stars evolve along the
AGB. Since we measure $^{12}$C/$^{13}$C below 20 we see four
possibilities to interpret this finding.  While we feel options
become increasing unlikely with position on the list we can not
exclude any of them on the basis of observations.

\noindent 
1. The carbon isotopic ratio was very low in these stars on their
arrival on the early AGB.  Carbon isotopic ratios as low as 10 $-$ 15
are observed for low mass RGB stars of near-solar metallicity
\citep{charbonnel_et_al98}.  Should the star experience thermal pulses without
or with only minor third dredge-up mixing 
\citep[e.g.,][]{lebzelter_hron03} a ratio between 10 and 20 is
compatible with theory.

\noindent 
2. Extra-mixing may occur on the AGB 
\citep{karakas_et_al_2010,busso_et_al_2010}.  \citet{lebzelter_et_al08}
brought up this process in order to interpret the unexpectedly low
$^{12}$C/$^{13}$C ratios observed in C-stars in the LMC cluster NGC
1846. The AGB stars in this cluster are thought to be of similar
mass, i.e., around 1.5\,$M_\odot$,
to the field Miras but with different metallicity.

\noindent
3. Variability studies demonstrating that Mira variability is limited
to the final AGB stages are only for the 
Magellanic Clouds and globular clusters 
\citep[see for instance][]{wood_2015}. 
Magellanic Clouds and globular clusters 
stars typically have sub-solar
metallicity.  Until the Gaia results are compiled the luminosities
of Galactic field Miras remain poorly known. It is possible that
at higher metallicity Mira variability is not limited to stars near
the termination of the AGB.

\noindent 
4. Some of the sample stars could be massive AGBs in which the HBB
regulates the carbon isotopic ratio toward lower values.  In this
case, we should have measured oxygen
isotopic ratios typical for this mass range. Actually, a few M stars in
our sample have $^{16}$O/$^{18}$O$>1000$ but only one, R Ser, also
has $^{12}$C/$^{13}$C\textless20. It should be noted that in $^{16}$O/$^{18}$O
measurement of R Ser is based on a single line.  In addition, the
period of R Ser is not as large as that expected for a massive Mira
star.

We note the presence of six stars in our sample with
$^{16}$O/$^{17}$O\textless200 but only one (RU Her) that shows a large
$^{16}$O/$^{18}$O.  In this case, however, the HBB hypothesis can
be discarded because the predicted HBB carbon isotopic ratio is $\sim5$
while the measure value is $25\pm5$.  As discussed by
\citet{lebzelter_et_al15} the most obvious explanation for peculiar
values of the oxygen isotopic ratio is that these are fossil records
of an anomalous composition, 
possibly a $^{16}$O depleted environment,
where these stars were formed.  In conclusion, we do
not find clear evidence for stars undergoing HBB.

\subsection{S and C stars: third dredge-up imprints}

Twelve LPVs in our sample have $^{12}$C/$^{13}$C ratios above 30.
In this group, four objects are classified as S stars and 
three as C stars.  Assuming that the abundances are intrinsic these LPVs
have experienced TDU events.  For six stars of this group observations
of Tc in their spectra
\citep{little_et_al87,lebzelter_hron03,uttenthaler_et_al11} confirm
the TDU hypothesis.  The mass range of this subset of stars should
be $1.5~M_\odot$~\textless~$M$~\textless~3~$M_\odot$.

There are two C-stars in our sample that have a $^{12}$C/$^{13}$C
ratio clearly incompatible with the TDU hypothesis, VX And and U
Cyg.
As discussed in the previous section an intrinsic C star
should have $^{12}$C/$^{13}$C$>25$ if its initial mass is between
$1.5 M_\odot$ and $2 M_\odot$ or $>70$ if the initial mass is
larger than 2 $M_\odot$. VX And and U Cyg have carbon
isotopic ratios clearly smaller than 25, $13\pm 5$ and $14\pm 6$,
respectively.  They are likely J- or R-type C stars with an initial
mass \textless1.5 $M_\odot$.  For VX And this conclusion is also supported
by its large $^{16}$O/$^{17}$O$, 1349\pm 500$, a value that fits
expectations for an AGB star with $M\sim1.3$ $M_\odot$.  
Indeed \citet{barnbaum_et_al_1996} classifies VX And as 
J4.5 or MS5.  U Cyg does not present as clear a case. 

U Cyg is classified by \citet{sanford_1950} as R8 and Np. 
\citet{barnbaum_1994} classifies it as Nep.  
\citet{keenan_1957} notes the presence of a weak Li 6708 line. 
U Cyg shows a very low $^{16}$O/$^{17}$O,
$158\pm30$, that is outside the range of any reasonable
nucleosynthesis scenario.  In principle, such a low $^{16}$O/$^{17}$O
may be due to deep extra-mixing that occurred during the RGB and/or
the AGB.  However, in that case $^{12}$C/$^{13}$C should be $\sim
5$ and $^{16}$O/$^{18}$O$~>~10^5$ (see Figure \ref{profile_2p0} and
\ref{tab_mix}).

The $^{12}$C/$^{13}$C for UX Dra is $28\pm 7$.  This is marginally
compatible with a TDU scenario.  Both the $^{12}$C/$^{13}$C and the
$^{16}$O/$^{17}$O, $631\pm 250$, are compatible  with a newborn
C-star with a mass of about 1.6 $\pm 0.1$ $M_\odot$ and C/O$\sim
1$.  
This scenario is also supported by the s-process overabundances observed by 
\citet{abia_et_al01}.
However, the short period of 168 days and the 
semi-regular variability type are more appropriate for an early-AGB star.

The other three carbon stars in the sample, S Cep, V Hya, and R
Lep, are clearly well above the TDU limit of 25 for $^{12}$C/$^{13}$C.
S Cep has long period of 487 days and a large $^{12}$C/$^{13}$C$, 224\pm130$,
both compatible with an evolved AGB star with $M\le 2$ $M_\odot$.
Similarly long period, 529 days, and high $^{12}$C/$^{13}$C, 
$200\pm 50$, are found for V Hya.  With these features V Hya is
probably an intrinsic C star of $2.5 \pm 0.5$ $M_\odot$ approaching
the AGB tip.  However, although affected by a large error the
$^{16}$O/$^{17}$O, $1122\pm 400$, is probably too high.  V Hya is
a peculiar LPV with high velocity flows and has been suggested as
a nascent proto-PN \citep{sahai_et_al_2009}.  R Lep has a long
period of 432 days coupled to $^{12}$C/$^{13}$C~$=$~$34\pm 5$ and
$^{16}$O/$^{17}$O~$=$~$676\pm 120$ in accord with expectations for an
evolved low-mass AGB star of $M\sim 1.6$ $M_\odot$ close to the AGB
tip.

Finally RZ Per, the single SC star in our sample, has a low
$^{12}$C/$^{13}$C of $9\pm 3$. SC stars are intriguing objects whose
nature has not been unveiled yet \citep{hedrosa13}.  Due to the
complex spectrum the oxygen isotopic ratios could not be derived
for this star.

\subsection{O-rich presolar grains from AGB stars.}

\begin{figure}
\epsscale{0.8}
\plotone{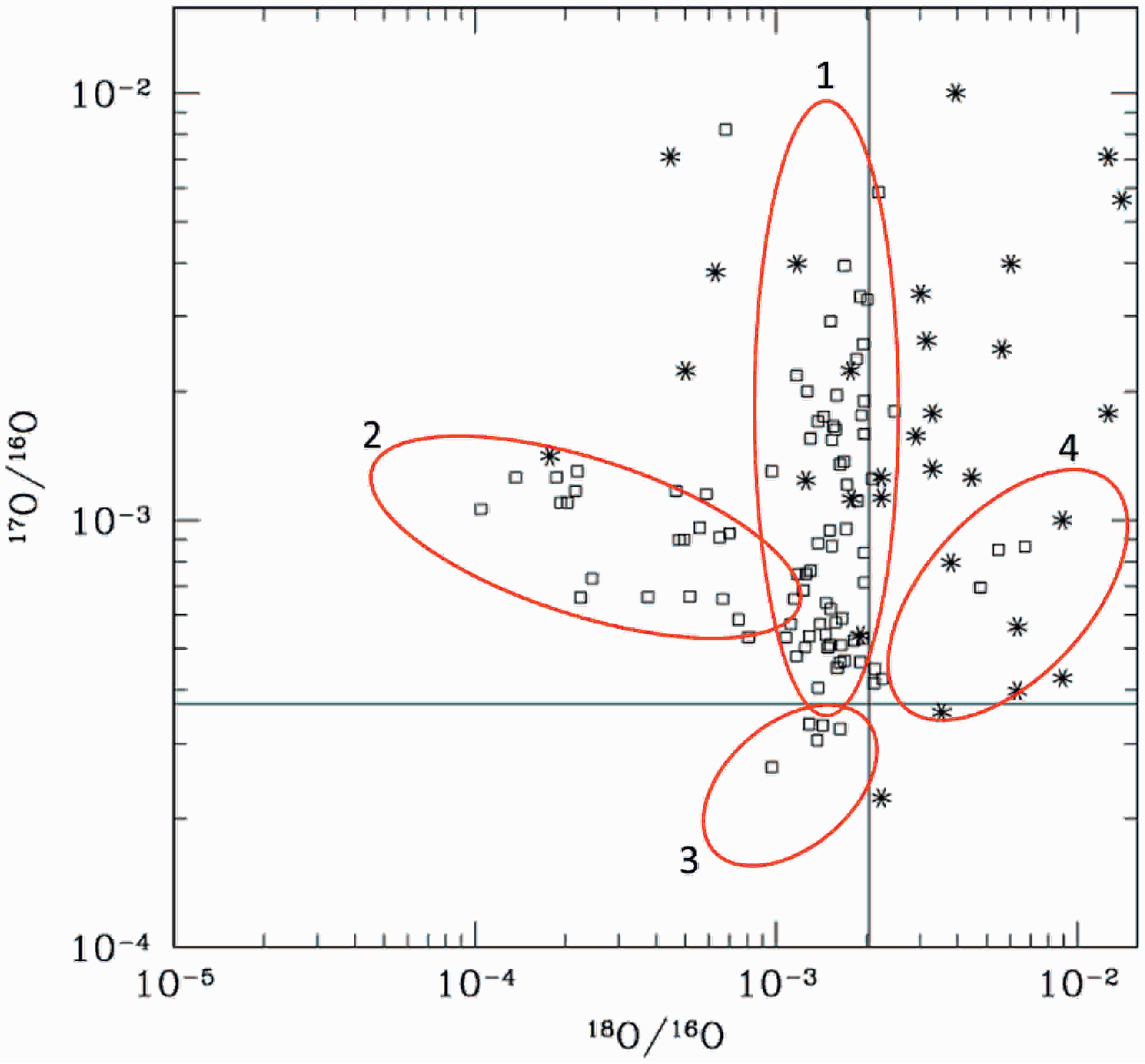}
\caption{
Open symbols, abundances measured in O-rich presolar grains from Table 2
of \citet{nittler_et_al_2008}. Starred symbols,  Mira measurements,
this paper.  The light solid lines show the solar values for the two oxygen
isotopes while the ellipses delimit regions occupied by different groups of grains (see text).
\label{fig14}}
\end{figure}

Presolar grains\footnote{In the meteoritic abundance community the
isotopic ratios are the inverse of the astronomical convention.  In
this section, we adopt their convention to allow comparison with
literature data.}, i.e. dust grains formed in the material ejected
by stars of different mass and composition and found in pristine
meteorites, bring clear signatures of the nucleosynthesis and mixing
processes occurred in the parent stars \citep{zinner_1998}.  Among
them the majority of the O-rich grains, oxides and silicates, are
believed to originate in the wind of AGB stars \citep{nittler_2009}.
Therefore, the isotopic abundances measured in these presolar
material can be directly compared to the corresponding abundances
we measured in O-rich LPVs. In Figure \ref{fig14} oxygen isotopic ratios
in presolar grains (squares) are compared to the values found for
the O-rich AGB stars in our sample.  The lifetime of O-rich grains
in the interstellar medium is $\sim 500$ Myr \citep{jones_1996}.
Thus the parent stars of the presolar grains should have ended their
life within a few hundreds Myr before the formation of the solar
system.  If the stars were low-mass AGBs then the time of their
formation coincides with a very early epoch of the Milky Way
evolution. On the contrary, the AGB stars in our sample were formed
no more than a couple of Gyr ago and, in any case, much later than
the birth of the solar system. This occurrence likely implies a
difference in the initial chemical composition: the present day
generation of AGB stars is, on the average, more metal-rich than
the AGB stars from which the presolar grains formed.

Keeping in mind this remark and following \citet{nittler_2009} we
note that the majority of the O-rich grains, those belonging to
group 1 in Figure \ref{fig14}, should come from low-mass AGB stars
with 1.2~\textless~$M/M_\odot$~\textless~2. As stressed in the previous sections,
the O isotopic ratios in these stars are determined by i) the initial
composition and ii) the first dredge up.  The quoted mass range is,
in fact, compatible with the spread of $^{17}$O/$^{16}$O resulting
after the first dredge up.  The sub-solar $^{18}$O/$^{16}$O of group
1 grains also supports such an hypothesis.  Indeed, it is compatible
with the $^{18}$O depletion caused by the first dredge up on material
whose original composition was nearly solar. The spread of
$^{17}$O/$^{16}$O observed in the stars of our sample is almost
identical to that of the O-rich presolar grains confirming that
they are AGB stars with mass similar to those from which the grains
originated. However, at variance with the grain abundances the
majority of the AGB stars in our sample present super-solar
$^{18}$O/$^{16}$O. This occurrence must reflect a difference in the
initial composition, namely a $^{16}$O depletion and/or a $^{18}$O
enhancement.  This is not surprising considering the different
formation epoch. Indeed, chemical evolution models predict a steadily
increase with time of the  $^{18}$O/$^{16}$O \citep{prantzos_et_al_1996}.
A similar conclusion was obtained by \citet{lebzelter_et_al15} by
studying the $^{18}$O/$^{16}$O ratios of giant stars in young open
clusters.

Grains belonging to group 2, those showing rather low $^{18}$O/$^{16}$O,
probably originated in stars with masses similar to those of group
1 but they must have suffered an additional $^{18}$O depletion
likely caused by some extramixing that took place during the AGB phase
\citep{nittler_et_al_2008}. Note that only one star in our sample,
SV Cas, shares O isotopic ratios similar to those of group 2
grains. Perhaps this occurrence indicates that AGB extramixing
is rarer in more metal-rich AGB stars.

The few grains belonging to groups 3 and 4 are believed to have a different origin, interstellar
medium for group 3 and supernova ejecta for group 4. 
Note, however, that some Miras of our sample show O isotopic ratios similar to those of group 4 grains.

As noted by \citet{nittler_2009}, no presolar grains from massive AGB, M$> 4$ M$_\odot$, have been found so far. 
Interestingly, we did not found Miras showing the chemical peculiarities caused by
the HBB, a feature expected for massive AGB stars.

\section{CONCLUSIONS}

We have presented $^{12}$C/$^{13}$C, $^{16}$O/$^{17}$O, and
$^{16}$O/$^{18}$O isotopic ratios for a sample of 46 long period
AGB Mira and SRa variables.  The sample was selected mainly by apparent
infrared brightness and represents the population I field near the
sun.  Measurement of the $^{12}$C/$^{13}$C and $^{16}$O/$^{17}$O
ratios allow the main sequence mass for these
field stars to be categorized.  Among the AGB stars we have studied:

\begin{itemize} 

\item the majority of the M stars have masses M~\textless~2~$M_\odot$;  

\item all three S stars are the result of the occurrence of third dredge-up
episodes;  

\item three of the C stars are compatible with the TDU hypothesis and another C star is
likely undergoing TDU;

\item two C stars are either J or peculiar N-type resulting from binary mass 
transfer or merger;

\item no clear evidence of HBB has been found in any of our sample stars.  
\end{itemize}

The overall abundance changes are in accord with previous analysis of O-rich presolar grains and
reinforce the hypothesis about the low-mass AGB origin of this stardust.

\acknowledgments 
The work of TL has been supported by the Austrian Science Fund FWF
under project number P23737-N16.  The National Optical Astronomy
Observatory is operated by the Association of Universities for
Research in Astronomy under cooperative agreement with the National
Science Foundation.  This research was facilitated by 
the SIMBAD database, operated by CDS in Strasbourg, France,
and NASA's Astrophysics Data System Abstract Service.  We acknowledge
with thanks the variable star observations from the AAVSO International
Database contributed by observers worldwide and used in this research.


\clearpage

\begin{deluxetable}{lclrccr}
\tabletypesize{\footnotesize}
\tablecaption{Observations \label{t:obslog}}
\tablewidth{0pt}
\tablehead{\colhead{Star} & \colhead{Type} & \colhead{Date} & \colhead{Int. Time} & \colhead{Apodizing} & \colhead{Resolution}  & \colhead{S/N}\\
\colhead{ } & \colhead{ } & \colhead{} & \colhead{(min.)} & \colhead{Fcn\tablenotemark{a}} & \colhead{(cm$^{-1}$)}  & \colhead{(peak)}
}
\startdata
R And        & M   & 78 Oct 17  &  68  &  2  &  0.074  &  114\\ 
VX And       & SRa & 79 Jul 18  & 135  &  2  &  0.070  &   69\\ 
R Aql        & M   & 77 Aug 26  &  60  &  2  &  0.084  &   98\\ 
R Aqr        & M   & 83 Sep 14  &  17  &  2  &  0.070  &  160\\ 
T Ari        & M   & 83 Jan 22  &  94  &  2  &  0.070  &   60\\ 
T Ari        & M   & 85 Jul 2   &  52  &  2  &  0.070  &   51\\ 
R Aur        & M   & 84 Aug 8   &  34  &  2  &  0.070  &   80\\ 
T Cam        & M   & 83 Jun 4   &  42  &  2  &  0.070  &   30\\
\enddata
\tablenotetext{a}{\citet{norton_beer_1976}}
\tablenotetext{b}{No reverse scan}
\tablenotetext{c}{Reference spectrum}
\tablecomments{The variability class is taken from the GCVS.}
\end{deluxetable}

\clearpage

\begin{deluxetable}{lccccrrr}
\tabletypesize{\footnotesize}
\tablewidth{0pt}
\tablecaption{Isotopic ratios.\label{t:results}}
\tablehead{
\colhead{Star} & \colhead{Type} & \colhead{Period [d]} & \colhead{Phase} & \colhead{T$_{\rm excit}$ [K]} & 
    \colhead{$^{12}$C/$^{13}$C} & \colhead{$^{16}$O/$^{17}$O} & \colhead{$^{16}$O/$^{18}$O}}
\startdata
R And     & S-Mira & 409 & 0.59 & 2500 & 40$\pm$15 & 1318$\pm$300 & 708$\pm$150\\
VX And   & C-SRa & 369 & 0.41: & 3200 & 13$\pm$5 & 1349$\pm$500 & 100$\pm$50\\ 
R Aqr      & Mira & 387 & 0.56: & 3100 & 16$\pm$5 & 2344$\pm$1000 & 112$\pm$40\tablenotemark{a}\\
R Aql      & Mira & 293 & 0.69 & 3100 & 8: & 224: & 398:\tablenotemark{a}\\ 
T Ari       & Mira & 323 & 0.4/0.71 & 2900/3200 & 15$\pm$5 & 807$\pm$450 & 795$\pm$230\\
R Aur      & Mira & 458 & 0.13 & 2950 & 33$\pm$13 & 562$\pm$300 & 302$\pm$150\\
T Cam    & S-Mira & 374 & 0.52 & 2600 & 31$\pm$10 & 158$\pm$70 & 219$\pm$90\\
TX Cam  & Mira & 557 & 0.52 & 2800 & 21$\pm$6 & 380$\pm$120 & 316$\pm$120\\
\enddata
\tablenotetext{a}{Ratio $^{16}$O/$^{18}$O based on a single line.}
\tablecomments{See \ref{symerr} for details on the uncertainty given in the Table.}
\end{deluxetable}

\clearpage

\begin{deluxetable}{lrrrrrrrrr}
\tabletypesize{\footnotesize}
\tablecaption{Comparison with literature values.\label{t:complit}}
\tablehead{
\colhead{Star} & \multicolumn{3}{c}{$^{12}$C/$^{13}$C} & \multicolumn{3}{c}{$^{16}$O/$^{17}$O} & 
  \multicolumn{3}{c}{$^{16}$O/$^{18}$O}\\
  \colhead{ } & \colhead{Lit.} & \colhead{Ref.} & \colhead{this} & \colhead{Lit.} & \colhead{Ref.} &   
  \colhead{this} & \colhead{Lit.} & \colhead{Ref.} & \colhead{this}\\
  \colhead{} & \colhead{value} & \colhead{ } & \colhead{paper} & \colhead{value} & \colhead{ } & \colhead{paper} 
  & \colhead{value} & \colhead{ } & \colhead{paper}}
\startdata
R And & 24        & (2) & 40$\pm$15& & & & & & \\
      & 33        & (7) &          & & & & & & \\
VX And& 13$\pm$3  & (1) & 13$\pm$5 & 400$\pm$175 & (4) & 1349$\pm$500 & $>$150 & (4) & 100$\pm$50\\
      & 6         & (13)&          & & & & & & \\
TX Cam& 15        & (2) & 21$\pm$6 & & & & & & \\
R Cas & 9         & (2) & 12$\pm$4 & & & & & & \\
$\chi$ Cyg & 40   & (2) & 35$\pm$4 & & & & & & \\
      & 39        & (7) &          & & & & & & \\
\enddata
\tablerefs{(1) \citet{lambert_et_al86}; (2) \citet{ramstedt_olofsson14}; (3) \citet{schoier_olofsson00};
(4) \citet{harris_et_al87}; (5) \citet{khouri_et_al14a}; (6) \citet{khouri_et_al14b};
(7) \citet{wallerstein_et_al11}; (8) \citet{dominy_wallerstein87}; (9) \citet{harris_lambert84};
(10) \citet{tsuji08}; (11) \citet{hedrosa13}; (12) \citet{abia_et_al01}; (13) \citet{abia_isern00}}
\tablecomments{Where literature values are listed without an uncertainty it was not possible to deduce the uncertainty
from the paper unambiguously. Throughout this paper, the uncertainties given are symmetric (see comment in 
\ref{symerr}). In the case that asymmetric uncertainties are given in the literature mean values are given here. For R Cyg, (8) states
an uncertainty of +2000 and $-$600 so in this case a mean value made no sense.}
\end{deluxetable}

\begin{deluxetable}{rrrrrrrrr}
\tabletypesize{\footnotesize}
\tablecaption{Appendix 1 CO $\Delta$v=2 Best Lines\label{a:bestcodv2}}
\tablehead{
\colhead{Wavenumber} &  \colhead{Wavelength} & \multicolumn{2}{c}{Isotope} & \multicolumn{4}{c}{Transition} & \colhead{Number of } \\
\colhead{(vacuum cm$^{-1}$)} & \colhead{(air \AA)} & \colhead{C} & \colhead{O} &  \colhead{v'} & \colhead{v''} & \colhead{Branch} & \colhead{J''} & \colhead{spectra} }
\startdata
4001.726 &  24982.40 & 12 & 16 &  3 &  1 & P &  40 &   6 \\
4037.166 &  24763.09 & 12 & 17 &  3 &  1 & P &  26 &   4 \\
4037.753 &  24759.49 & 12 & 18 &  2 &  0 & P &  27 &   5 \\
4037.975 &  24758.13 & 12 & 16 &  6 &  4 & P &   3 &   5 \\
4049.647 &  24686.77 & 12 & 16 &  6 &  4 & R &  97 &   8 \\
4055.447 &  24651.47 & 12 & 16 &  4 &  2 & P &  22 &   7 \\
4055.641 &  24650.29 & 13 & 16 &  2 &  0 & P &  25 &   6 \\
4056.826 &  24643.09 & 12 & 16 &  5 &  3 & P &  11 &   4 \\
\enddata
\end{deluxetable}

\begin{deluxetable}{rrrrrrrrr}
\tabletypesize{\footnotesize}
\tablecaption{Appendix 2 CO $\Delta$v=3 Best Lines\label{a:bestcodv3}}
\tablehead{
\colhead{Wavenumber} &  \colhead{Wavelength} & \multicolumn{2}{c}{Isotope} & \multicolumn{4}{c}{Transition} & \colhead{Number of } \\
\colhead{(vacuum cm$^{-1}$)} & \colhead{(air \AA)} & \colhead{C} & \colhead{O} &  \colhead{v'} & \colhead{v''} & \colhead{Branch} & \colhead{J''} & \colhead{spectra} }
\startdata
5568.210 &  17954.19 & 12 & 16 & 10 &  7 & P &  41 &   4 \\
5587.946 &  17890.78 & 12 & 16 &  9 &  6 & P &  48 &   4 \\
5631.431 &  17752.63 & 12 & 16 & 11 &  8 & P &  20 &   4 \\
5638.522 &  17730.30 & 12 & 16 &  8 &  5 & P &  51 &   7 \\
5647.716 &  17701.44 & 12 & 16 & 11 &  8 & P &  17 &   6 \\
5669.443 &  17633.60 & 12 & 16 & 12 &  9 & R &   7 &   4 \\
5669.814 &  17632.45 & 12 & 16 &  7 &  4 & P &  56 &   7 \\
5675.918 &  17613.49 & 12 & 16 & 12 &  9 & R &  53 &   6 \\
\enddata
\end{deluxetable}

\end{document}